\documentclass[]{aastex631}

\shorttitle{Characterization of of (98943) 2001 CC$_{21}$, the target of Hayabusa2$\#$ }
\shortauthors{Marcel Popescu, Eri Tatsumi, et al.}
\graphicspath{{./}{figures/}}
\begin{document}

\title{Ground-based characterization of (98943) Torifune  2001 CC$_{21}$, the target of Hayabusa2$\#$ space mission\footnote{Based on observations made with the the Two-meter Twin Telescope (TTT) and the Telescopio Carlos Sanchez (TCS) in the Spanish Observatorio del  Teide, Gran Telescopio Canarias (GTC) telescope and the Isaac Newton Telescope (INT) in the Spanish Observatorio del Roque de los Muchachos of the Instituto de Astrof\'{\i}sica de Canarias, the NASA Infrared Telescope Facility (NASA IRTF) at the Mauna Kea Observatory in Hawaii.}}

\author[0000-0001-8585-204X]{Marcel M. Popescu}
%\altaffiliation{These authors contributed equally:  M. Popescu (popescu.marcel@ucv.ro), E. Tatsumi (tatsumi.eri@jaxa.jp).}
\affiliation{University of Craiova \\
Str. A. I. Cuza nr. 13, 200585 Craiova, Romania}
\affiliation{Astronomical Institute of the Romanian Academy \\
5 Cu\c{t}itul de Argint, 040557 Bucharest, Romania}
\email{popescu.marcel@ucv.ro}

\author[0000-0002-6142-9842]{Eri Tatsumi}
%\altaffiliation{These authors contributed equally:  M. Popescu (popescu.marcel@ucv.ro), E. Tatsumi (tatsumi.eri@jaxa.jp).}
\affiliation{Institute of Space and Astronautical Science (ISAS) \\
Japan Aerospace Exploration Agency (JAXA) \\
Sagamihara, 252-5210 Kanagawa, Japan}
\email{tatsumi.eri@jaxa.jp}

\author[0000-0002-9214-337X]{Javier Licandro}
\affiliation{Instituto de Astrof\'{\i}sica de Canarias (IAC) \\
C/V\'{\i}a L\'{a}ctea s/n, 38205 La Laguna, Tenerife, Spain}
\affiliation{Departamento de Astrof\'{\i}sica \\
Universidad de La Laguna, 38206 La Laguna, Tenerife, Spain}

\author[0000-0002-8134-2592]{Miguel R. Alarcon}
\affiliation{Instituto de Astrof\'{\i}sica de Canarias (IAC) \\
C/V\'{\i}a L\'{a}ctea s/n, 38205 La Laguna, Tenerife, Spain}
\affiliation{Departamento de Astrof\'{\i}sica \\
Universidad de La Laguna, 38206 La Laguna, Tenerife, Spain}

\author[0000-0002-2718-2022]{Javier Rodr\'{\i}guez Rodr\'{\i}guez}
\affiliation{Instituto Universitario de Ciencias y Tecnolog\'{\i}as Espaciales de Asturias (ICTEA), University of Oviedo, \\
C. Independencia 13, E-33004 Oviedo, Spain}

\author[0000-0002-2394-0711]{Miquel Serra-Ricart}
\affiliation{Instituto de Astrof\'{\i}sica de Canarias (IAC) \\
C/V\'{\i}a L\'{a}ctea s/n, 38205 La Laguna, Tenerife, Spain}
\affiliation{Departamento de Astrof\'{\i}sica \\
Universidad de La Laguna, 38206 La Laguna, Tenerife, Spain}
\affiliation{Light Bridges S. L., Observatorio del Teide, Carretera del Observatorio S/N, E-38500 Guimar, Tenerife, Canarias, Spain}

\author[0000-0002-0696-0411]{Julia de Le\'on}
\affiliation{Instituto de Astrof\'{\i}sica de Canarias (IAC) \\
C/V\'{\i}a L\'{a}ctea s/n, 38205 La Laguna, Tenerife, Spain}
\affiliation{Departamento de Astrof\'{\i}sica \\
Universidad de La Laguna, 38206 La Laguna, Tenerife, Spain}

\author{Joaqu\'in Fernandez Martin}
\affiliation{Instituto de Astrof\'{\i}sica de Canarias (IAC) \\
C/V\'{\i}a L\'{a}ctea s/n, 38205 La Laguna, Tenerife, Spain}
\affiliation{Astronomical Institute of the Romanian Academy \\
5 Cu\c{t}itul de Argint, 040557 Bucharest, Romania}

\author{David Morate}
\affiliation{Centro de Estudios de F\'{\i}sica del Cosmos de Arag\'{\o}n (CEFCA)\\
Unidad Asociada al CSIC, Plaza San Juan 1, 44001}

\author{Gabriel N. Simion}
\affiliation{Astronomical Institute of the Romanian Academy \\
5 Cu\c{t}itul de Argint, 040557 Bucharest, Romania}

\author[0000-0003-3166-5139]{Bogdan Alexandru Dumitru}
\affiliation{Institute of Space Science (ISS) \\
409, Atomi\c{s}tilor Street, 077125 M\u{a}gurele, Ilfov, Romania}

\author{Daniel Nicolae Berte\c{s}teanu}
\affiliation{Astroclubul Bucure\c{s}ti \\
Blvd Lasc\u{a}r Catargiu 21, 10663 Bucharest, Romania}
\affiliation{Astronomical Institute of the Romanian Academy \\
5 Cu\c{t}itul de Argint, 040557 Bucharest, Romania}

\author{George Pantelimon Prodan}
\affiliation{University of Craiova \\
Str. A. I. Cuza nr. 13, 200585 Craiova, Romania}
\affiliation{Astronomical Institute of the Romanian Academy \\
5 Cu\c{t}itul de Argint, 040557 Bucharest, Romania}

\author[0000-0002-1821-5689]{Masatoshi Hirabayashi}
\affiliation{Georgia Institute of Technology, Atlanta, GA 30320, United States}

%% Note that the \and command from previous versions of AASTeX is now
%% depreciated in this version as it is no longer necessary. AASTeX 
%% automatically takes care of all commas and "and"s between authors names.

%% AASTeX 6.31 has the new \collaboration and \nocollaboration commands to
%% provide the collaboration status of a group of authors. These commands 
%% can be used either before or after the list of corresponding authors. The
%% argument for \collaboration is the collaboration identifier. Authors are
%% encouraged to surround collaboration identifiers with ()s. The 
%% \nocollaboration command takes no argument and exists to indicate that
%% the nearby authors are not part of surrounding collaborations.

%% Mark off the abstract in the ``abstract'' environment. 
\begin{abstract}
The near-Earth asteroid (98943) Torifune, previously designated 2001 CC$_{21}$, is the flyby target of the Hayabusa2 extended mission, nicknamed Hayabusa2$\#$ (SHARP: Small Hazardous Asteroid Reconnaissance Probe). The ground-based telescope observations offer a key science input for the mission's scientific investigation. During 2022 - 2024  this asteroid was at visible apparent magnitudes brighter than 18.5, allowing for a detailed characterization using ground-based telescope observations.  We determined its rotation period $P~=~5.021516\pm0.000106$ h and its absolute magnitude H = 18.78 $\pm$ 0.14 and.  The large number of lightcurves allows to estimate its axes ratio, its convex shape and its pole orientation $\lambda = 301^{\circ} \pm 35^{\circ}$, $\beta = {89^{+1}_{-6}}^{\circ}$ and $\epsilon = 5^{\circ} \pm 3^{\circ}$ which indicate a prograde rotation.  We report the semi-axis of the equivalent ellipsoid, $a$ = 0.42$^{+0.08}_{0.06}$ km, $b$ = 0.16$^{+0.05}_{0.04}$ km,  and $c$ = $0.17\pm0.03$ km.  Consequently, the volume equivalent diameter is $D_{eq}$ = $0.44 \pm 0.06$ km . Using observations conducted simultaneously with four broadband filters, we determined $(g-r) = 0.663 \pm 0.022$ mag, $(r-i) = 0.177 \pm 0.012$ mag, and $(i-z_s) = -0.061 \pm 0.032$ mag. Additionally, we found that Torifune exhibits no detectable large-scale heterogeneity. We classified the object using a high signal-to-noise ratio spectrum (over the visible and near-infrared region) as Sq-type in the Bus-DeMeo taxonomy.  We estimate a mineralogy similar to LL/L ordinary chondrites, with an ol/(ol+px) = 0.60, a Fa content of 28.5 mol$\%$, and a Fs content of 23.4 mol$\%$. The spectral data indicate a surface affected by moderate space weathering effects.

\end{abstract}

%% Keywords should appear after the \end{abstract} command. 
%% The AAS Journals now uses Unified Astronomy Thesaurus concepts:
%% https://astrothesaurus.org
%% You will be asked to selected these concepts during the submission process
%% but this old "keyword" functionality is maintained in case authors want
%% to include these concepts in their preprints.
\keywords{Minor planets, asteroids: individual: (98943) Torifune 2001 CC$_{21}$ --
             -- techniques: spectroscopic -- techniques: photometric -- 
             methods: observational}

%% From the front matter, we move on to the body of the paper.
%% Sections are demarcated by \section and \subsection, respectively.
%% Observe the use of the LaTeX \label
%% command after the \subsection to give a symbolic KEY to the
%% subsection for cross-referencing in a \ref command.
%% You can use LaTeX's \ref and \label commands to keep track of
%% cross-references to sections, equations, tables, and figures.
%% That way, if you change the order of any elements, LaTeX will
%% automatically renumber them.
%%
%% We recommend that authors also use the natbib \citep
%% and \citet commands to identify citations.  The citations are
%% tied to the reference list via symbolic KEYs. The KEY corresponds
%% to the KEY in the \bibitem in the reference list below. 

\section{Introduction} \label{sec:intro}

The Japanese asteroid sample-return mission Hayabusa2 completed its nominal operations of investigating and returning samples from the carbonaceous asteroid (162173) Ryugu in December 2020 \citep[e.g.][]{2022Sci...375.1011T}. The spacecraft continues its voyage as an extended mission \citep{2021AdSpR..68.1533H}, nicknamed Hayabusa2\# (SHARP -- Small Hazardous Asteroid Reconnaissance Probe). It aims to rendezvous with the fast-rotating near-Earth asteroid 1998 KY$_{26}$ \citep{1999Sci...285..557O} in 2031.

On the way to 1998 KY$_{26}$, Hayabusa2\# will flyby the asteroid (98943) Torifune, previously designated 2001 CC$_{21}$, in 2026. During the closest approach (CA), the spacecraft will flyby the body as closely as possible, probably less than 10 km. The Optical Navigation Camera Telescope (ONC-T) can acquire images with a resolution of about $\approx$10~m/pixel if the CA distance is about 100 km, and the Thermal-Infrared Camera (TIR) can get data with a sampling of $\approx$90~m/pixel. The near-infrared spectrometer (NIRS3) will observe the global and local spectra raging from 1.8 to 3.2 $\mu m$. These measurements will allow us to observe the global shape, the largest features (craters and boulders) on the surface, the overall thermal heterogeneity, and the surface visible-to-near-infrared spectral properties \citep{2021AdSpR..68.1533H}. In this context, the observations performed with the ground-based telescopes are a key input for planning the scientific investigation of this object. Especially, there is a wavelength gap between ONC and NIRS3 which can be filled using telescopic observations with a high signal-to-noise ratio. These telescope measurements are also important for designing the engineering approach of the spacecraft to this near-Earth object to maximize the scientific return. 
%
%---------------------------------------
%--------- TABLE I rbital elements 98943
   \begin{table}
    \centering
       \fontsize{8}{12pt}\selectfont
       \tabcolsep 0.14truecm
       \caption{\label{elements}Values of the heliocentric Keplerian orbital elements and their respective 1$\sigma$ uncertainties of (98943) Torifune. The orbit determination of (98943) Torifune is referred to epoch  2460400.5 (31 March 2024) TDB (Barycentric Dynamical Time, J2000.0 ecliptic and equinox) and it is based on 1393 observations with a data-arc span of 15083 days, 41.30 years (solution date, 2024-Nov-12 08:25:51). Source: JPL SBDB.
               }
       \begin{tabular}{lcc}
        \hline
         Orbital parameter                                 &   & value$\pm$1$\sigma$ uncertainty \\
        \hline
         Semi-major axis, $a$ (au)                         & = &   1.0323128936744$\pm$3.8134E-9 \\
         Eccentricity, $e$                                 & = &   0.2192370700348133$\pm$2.2204E-8\\
         Inclination, $i$ (\degr)                          & = &   4.807777191687606$\pm$1.5085E-6  \\
         Longitude of the ascending node, $\Omega$ (\degr) & = &  75.44790934304839$\pm$1.5831E-5  \\
         Argument of perihelion, $\omega$ (\degr)          & = & 179.3988630124814$\pm$1.6745E-5   \\
         Mean anomaly, $M$ (\degr)                         & = & 124.7628327227654$\pm$8.1552E-6   \\
         Perihelion distance, $q$ (au)                     & = &   0.8059916395060651$\pm$2.2865E-8  \\
         Aphelion distance, $Q$ (au)                       & = &   1.258634147842736$\pm$4.6494E-9 \\
         Absolute magnitude, $H$ (mag)                     & = &  18.73                            \\
         Earth MOID (au) 	                              & = &   0.0829663                    \\
        \hline
       \end{tabular}
   \end{table}
%
%---------------------------------------

Torifune is a near-Earth asteroid (NEA) with an Apollo-type orbit, it has the aphelion at  Q = 1.26 au and the perihelion inside the Earth orbit at q = 0.81 au (the exact values and the associate uncertainties are shown in Table~\ref{elements}). Thus, it makes close approaches to Earth with a minimum orbital intersection distance (MOID) of 0.08 au. 

In the last decades, there were several favorable observing apparitions for Torifune  (when asteroids are at an apparent V magnitude brighter than $\approx$18, they are accessible to small and medium class telescopes, and various observing techniques can be applied). After its discovery, the object made four Earth close approaches during the years 2001, 2002, 2004, and 2023\footnote{\url{https://newton.spacedys.com/neodys/index.php?pc=1.1.8&n=98943}}. The information about the physical properties of this NEA could be determined using photometric, spectroscopic, and polarimetric observations. The next favorable chance for obtaining new data about Torifune will be in August 2025. 

A preliminary solution for the spin state and the shape was presented by \citet{FatkaPGC}. They used lightcurves obtained during  2002, 2003, 2022, and 2023, with the 1.54~m Danish telescope located in La Silla (Chile), Mayer 0.65-m telescope at Ond{\v{r}}ejov (Czechia), and from Asteroid Lightcurve Data Exchange Format (ALCDEF) Database\footnote{\url{https://alcdef.org/index.php}}. \citet{FatkaPGC} report a period P = 5.022 $\pm$ 0.001 h, and their convex shape model shows an elongated and flattened figure.

The visible spectrum of Torifune was first reported in 2004 by \citet{2004M&PS...39..351B}. This first observation classified the asteroid's reflectance spectrum as L-type (Bus taxonomy). It has a red slope and is relatively flat around 1 $\mu m$. The L-type is a peculiar class of asteroids \citep{2018Icar..304...31D}, and less than 3$\%$ among all spectrally observed NEAs belong to it \citep{2019Icar..324...41B}. Furthermore, \citet{2009Icar..202..160D} merged this spectrum with its near-infrared counterpart (0.8 -- 2.5 $\mu m$) observed on October 24, 2004 with SpeX instrument at the Mauna Kea 3.0 m NASA Infrared Telescope Facility (IRTF) and classify it as an Sw type (Bus - DeMeo taxonomy system).

During the same favorable observational apparition, \citet{2005MNRAS.359.1575L} reported a visible to near-infrared spectral curve of Torifune. They classified it as an Sk-type (using Bus taxonomy, which covers only the visible wavelengths). The significant difference compared with the data of \citet{2009Icar..202..160D} is the less red spectral slope and a stronger 1 $\mu m$ absorption band characteristic of olivine-pyroxene compositions. Mostly, these differences are due to the visible part.

Recently, spectral observations over the 0.7 -- 2.5 $\mu m$ wavelengths were reported by \citet{2023MNRAS.525L..17G}. They used the SpeX/IRTF instrument and found that Torifune's spectral features are consistent with the S complex type asteroids, namely the clear absorption around 0.9 $\mu m$ and a shallow absorption around 1.9 $\mu m$ associated with pyroxene. These authors did not identify any spectroscopic variation with respect to asteroid rotation.

\citet{2023MNRAS.525L..17G} also reported optical polarimetric observations. They determined a polarimetric inversion angle $\alpha ~ \approx ~ 20^\circ$ and estimated an albedo of $p_V$ = 0.23$~\pm~$0.04.%, and the regolith grain sizes in the range of 100 -- 130 $\mu m$.
They concluded that their overall findings are consistent with an S-type asteroid and contradict the expected properties of an L-type asteroid.

In this context, we aimed a thorough characterization of Torifune by acquiring telescopic data during its close approach to Earth between 2022 and 2024. During this time interval, the asteroid reached apparent magnitudes as bright as  16.5 mag in V band. Our objectives were: 1) to classify this asteroid based on high signal-to-noise (SNR) visible to infrared spectrum; 2) to determine its surface composition; 3) to characterize the space-weathering effects on its surface; 4) to check for possible large-scale heterogeneity; 5) to determine its rotational properties; and  6) to estimate its shape. The observations were performed using various telescopes from the Teide observatory and the Roque de los Muchachos observatory in the Canary Islands (including the 10.4-m Gran Telescopio Canarias -- GTC), and with the NASA 3.2-m Infrared Telescope Facility (IRTF). These data will allow us to directly compare the ground-based telescopic observations and the in-situ measurements that will be obtained by the onboard instruments of Hayabusa2 spacecraft, the multi-band imager ONC-T, and the near-infrared spectrometer NIRS3.

The article is organized as follows, Section 2 describes the instruments used to acquire data, the observations, and data reduction procedures. In Section 3, we present the results of this program. In Section 4, we compare our findings with those reported by other authors, and discuss the implications for the Hayabusa2$\#$  mission. Section 5 concludes the article with a summary of all of our results.

\section{Observations and data reduction} \label{sec:obs}

We used three observational modes to characterize Torifune. First, continuous imaging (photometric observations) traces the object's lightcurve. With these data we can compute the rotational state, the approximate shape, and the absolute magnitude. Most of these observations were obtained with the Two-meter Twin Telescope facility (TTT). Some additional data were acquired with the 80~cm  Instituto de Astrof{\'i}sica de Canarias Telescope (IAC80), and with the 0.25~m T025 - Big Data for Small Bodies telescope (T025-BD4SB).

Second, the spectro-photometric data (photometric observations obtained using different broadband filters to compute colors) provide information on taxonomic classification and enable the search for large-scale heterogeneity. These observations were acquired mostly with the 1.52 m Telescopio Carlos S\'anchez Telescope (TCS), and a few data were obtained with the 2.54 m Isaac Newton Telescope (INT) and with the TTT.

Third, spectral observations over the visible and near-infrared wavelengths allow us to classify the asteroid \citep{2022A&A...665A..26M}, estimate its surface composition \citep{2015aste.book...43R}, and some of space-weathering effects on the surface \citep{2015aste.book..597B}. These kinds of observations were performed with the 10.4 m Gran Telescopio Canarias (GTC), which acquired the visible spectrum, and with  the 3.0 m Infrared Telescope Facility (NASA/IRTF) for the near-infrared wavelengths.

The IAC80, TTT, and TCS telescopes are located at the  Teide Observatory (latitude: 28{\degr}~18'~01.8"~N; longitude: +16{\degr}~30'~39.2"~W; altitude:  2386.75~m), Minor Planet Center(MPC) observation code 954 (TTT have their own codes, Y65 for TTT1 and Y66 for TTT2), in Tenerife (Canary Islands, Spain). The INT and GTC are at the Observatorio del Roque de los Muchachos (MPC code 950) at an altitude of 2\,396\,m, in La Palma (Canary Islands, Spain). All these five telescopes are part of the Observatorios de Canarias complex. The NASA/IRTF telescope is mostly dedicated to infrared astronomy, and it is located at the Mauna Kea Observatory (MPC code 568) on the island of Hawai'i, USA. The T025-BD4SB (MPC code 073) is a small outreach telescope built from professional - amateur collaboration, which is located in Bucharest, Romania at the Astronomical Institute of the Romanian Academy. Below, we briefly describe the instruments used with these telescopes, the observation planning and circumstances, and the data reduction methods applied for each setup.

\subsection{The photometric data} \label{subsec:lightcurves}

The photometric observations were acquired during 51 different nights. A total number of 31\,637 exposures were obtained. 
The appendix Appendix Table \ref{photomlogtable} provides the log of the photometric observations.  These were acquired during 52 different nights. The phase angle of Torifune varied between 5$^\circ$ and 83$^\circ$ during this time frame. 

{\bf TTT.} The majority of the lightcurves presented in this paper were obtained with the TTT facility. With this facility, we made 49 nights of observations, which were split into two campaigns, one in February -  March 2023 and the other between November 2023 and January 2024. The two TTT telescopes (TTT1 and TTT2) have an aperture of 0.80~m (they are prototypes of the planned 2~m aperture ones) and the focal ratios of f/6.85 and f/4.4, respectively. TTTs are installed on alt-azimuth mounts.

The observations were performed using the broadband filters, the $Luminance$ covers the 0.4 to 0.7 $\mu$m wavelength interval, and the standard Sloan $g$,  $r$, $i$, and $z_s$ filter. The exposure time was dynamically set (between 10 and 60 sec) to ensure the target's signal-to-noise ratio (SNR) was higher than 50. A total of 15 observing runs were performed with TTT1, and 34 with TTT2. 

%Images were acquired using the QHY411M\footnote{\url{https://www.qhyccd.com/}} scientific complementary metal-oxide semiconductor (sCMOS) detector and the iKon-L 936 \footnote{\url{https://andor.oxinst.com/products/ikon-xl-and-ikon-large-ccd-series/ikon-l-936}} charge-coupled device (CCD)  installed on one of the Nasmyth ports of each telescope . The QHY411M sensors have 151 megapixels with a pixel size of 3.76~$\mu$m~pixel$^{-1}$. This provides an effective field-of-view (FoV) of 51.4$^{\prime}\times$38.3$^{\prime}$ (with an angular resolution of 0.22"~pixel$^{-1}$) in TTT1 and 33.1$^{\prime}\times$24.7$^{\prime}$ (angular resolution of 0.14"~pixel$^{-1}$) in TTT2. The iKon-L 936 CCD .... {\bf Miguel please help}

The images in TTT1 were acquired using the iKon-L 936 camera, with a BEX-DD back-illuminated charge couple device (CCD) sensor. It has $2048~\times~2048$ pixels with a size of 13.5$~\mu$m/pixel, resulting in an effective field of view (FoV) of $17.3^{\prime}~\times~17.3^{\prime}$  with a scale of 0.51 arcsec pixel$^{-1}$. In the case of TTT2 telescope the QHY411M camera was used, which has a back-illuminated scientific complementary metal-oxide semiconductor \citep{2023PASP..135e5001A}. It has 151 megapixels with a size of 3.76 $\mu$m pixel$^{-1}$. The effective FoV was $51.4^{\prime}~\times~38.3^{\prime}$ and the plate scale was 0.22 arcsec pixel$^{-1}$.

{\bf IAC80}. We used this telescope during four observing nights between November 2022 and March 2023. The IAC80 is a telescope with an aperture of 82~cm and a focal ratio of 11.3 in the Cassegrain focus. The images were acquired with the CAMeLOt-2 camera, a back-illuminated e2V CCD camera of 16 megapixels, with each pixel covering 15 $\mu {\rm m}^2$. The plate scale is 0.32~arcsec pixel$^{-1}$, and the FoV of 21.98$^{\prime}\times$22.06~$^{\prime}$.  The Sloan $r'$ broad-band filter was mounted in front of the camera.

{\bf The T025 - BD4SB}. During February 2023, the brightness of our target was estimated to be around 16.5 V magnitude.  This represented an opportunity to make photometric observations of Torifune with a small aperture telescope. Thus, as a professional-amateur educational project, we acquired data using the T025-BD4SB telescope together with amateur astronomers from Astroclubul Bucure\,{s}ti. This telescope is a 0.25 m aperture, with a focal ratio of f/4, a Newtonian telescope mounted on a robotized equatorial mount (Skywatcher EQ6-pro ). The images were obtained with the QHY294M\footnote{\url{https://www.qhyccd.com/astronomical-camera-qhy294/}} CMOS camera, in 11.7 megapixels mode, with a resolution of 0.956 arcsecpixel$^{-1}$. The exposure time was set to 45 sec or 60 sec, depending on the night. 

We used various software applications and libraries to apply the data reduction tasks, namely, the Image Reduction and Analysis Facility -- IRAF \citep{1986SPIE..627..733T} $apphot$ package, the Python -- $Astropy$ package \citep{astropy:2013, astropy:2018, astropy:2022}, the GNU Astronomy Utilities -- Gnuastro \citep{gnuastro,noisechisel_segment_2019}, GNU Octave \citep{octave}, and Tycho\footnote{\url{https://www.tycho-tracker.com/}}. In general, each dataset was reduced with two different pipelines. 

The pre-processing of the data consists of dark subtraction and flat-field correction. The calibration images were acquired at the beginning and end of the night. On the corrected images, the aperture photometry (with an inner circular disk, a mid-border ring, and an external sky-ring) was performed either with IRAF $apphot$, or with PHOTOMETRYPIPELINE (PP) --\citet{2017A&C....18...47M},  or with Tycho software. The apertures were optimized to maximize the SNR of the light curves.

TTT images were reduced and processed using their own pipeline, resulting on the final photometric files that were used in this paper, together with the other telescopes data and reduction methods. These TTT routines form a comprehensive package of Graphic Processing Unit (GPU) based algorithms that pre-process, astrometrize and obtain the photometric data of every point source in the images  (Alarcon et al. in preparation).Finally, we manually removed the outliers (values that are outside $\sim\pm3\sigma$ interval with respect to the neighbouring points in the lightcurve). 

We compared the different results obtained by the different pipelines (slight variations may occur due to the way the software computes the values inside the aperture or in the sky-ring). Nevertheless, as a benchmark for publishing the data, we took the lightcurves produced by the PP. This software package, written in Python, obtains the calibrated photometry from the FITS images by performing the astrometric registration, aperture photometry, photometric calibration, and asteroid identification. PP uses the \emph{Astromatic} suite\footnote{\url{https://www.astromatic.net/}}, namelly SExtractor for source identification and aperture photometry \citep{1996A&AS..117..393B}, SCAMP for astrometric calibration \citep{2006ASPC..351..112B}, and SWarp for image regridding and co-addition \citep{2002ASPC..281..228B}. It also uses the JPL Horizons\footnote{\url{https://ssd.jpl.nasa.gov/horizons/}} module for obtaining the Solar System objects ephemeris to identify them in the images. The GAIA catalog \citep{2018A&A...616A...1G,2018yCat.1345....0G} was used for astrometric registration.  The PanSTARRS catalog \citep{2012ApJ...750...99T, 2016arXiv161205560C} was used for photometric calibration of the data acquired with the Sloan filters, and APASS DR9 \cite{2016yCat.2336....0H} for the rest. No particular criterion was applied for the calibration stars.

PP uses the aperture photometry performed by the Source Extractor. It finds the optimum aperture radius based on a curve-of-growth analysis \citep{2000hccd.book.....H,2017A&C....18...47M}.  This is defined as the smallest aperture radius at which at least 70$\%$ of each of the total target flux and the total background flux is included, and at the same time, the difference between the normalized target and background flux levels is smaller than 5$\%$ \citep{2017A&C....18...47M}.

\begin{equation}
    Luminance_{\rm IKON}[mag] = 0.4687 * g + 0.5313 * r
    \label{LuminaceI}
\end{equation}
\begin{equation}
    Luminance_{\rm QHY}[mag]  = 0.5159 * g + 0.4840 * r
    \label{LuminaceQ}
\end{equation}
The $Luminance$ filters used on the TTT telescopes represent a particular case. Their magnitude equations are provided in Eqs.~\ref{LuminaceI} and \ref{LuminaceQ}, where $g$ and $r$ correspond to the magnitudes for these filters. Due to the differing quantum efficiency functions of the IKON and QHY cameras across wavelengths, the equations vary slightly. Furthermore, there is an offset of 0.173 magnitudes between the $Clear$ filter used on T025-BD4SB and the $Luminance_{\rm QHY}$.

\subsection{ The spectro - photometric data} \label{subsec:colors}
We used the 1.52~m Telescopio Carlos S\'anchez (TCS) equipped with the MuSCAT2  \citep{2019JATIS...5a5001N} instrument to obtain simultaneous photometric observations in four broad-band filters, namely $g$ (400 -- 550 nm), $r$ (550 -- 700 nm), $i$ (700 -- 820 nm), and $z_s$ (820 -- 920 nm). This optical system has four channels, each ending with an independently controllable CCD camera (1024 $\times$ 1024 pixels). Thus, for each detector, the resolution is of $\approx$0.435 "/pixel and a FoV of 7.4$^\prime$ $\times$ 7.4$^\prime$.

\begin{table}
\centering
\caption{The observations log for the TCS telescope. The UT time for the start of observations, the estimated apparent V magnitude, the heliocentric and geocentric distances $r$ and $\Delta$, the phase angle $\alpha$, the total observing time during that night $t_{obs}$, the number of images acquired for a channel $N_{img}$ and the available filters (because of some technical issues with one of the cameras, only three channels were functional during the first nights) are provided.}
\label{TCSlog}
\begin{tabular}{l c c c c c c c} \hline\hline
${\rm UT_{start}}$ & $V_{\rm mag}$ & $r$ [au] & $\Delta$ [au] & $\alpha$ [$^\circ$] & $t_{\rm obs}$ [h] & $N_{\rm img}$ &Filters\\
2022-02-08T03:21 & 17.9 & 1.03474105 & 0.24155973 & 71.8 & 0.370 & 37  & $g,r,z_s$\\
2022-02-17T04:03 & 18.0 & 1.00087257 & 0.22467978 & 80.3 & 2.649 & 141 & $g,r,z_s$\\
2022-02-24T05:04 & 18.2 & 0.97411956 & 0.22188561 & 87.5 & 0.766 & 50   & $g,r,z_s$\\
2022-12-23T00:15 & 18.3 & 1.22033652 & 0.32339779 & 37.5 & 6.807 & 692 & $g,r,i,z_s$\\
2023-02-20T21:58 & 16.6 & 1.05386269 & 0.13386069 & 57.7 & 0.519 & 79  & $g,r,i,z_s$\\
2023-12-15T23:03 & 17.1 & 1.25179787 & 0.27337324 & 10.6 & 6.586 & 656 & $g,r,i,z_s$\\
\hline

\end{tabular}
\end{table}
% TCS DATA status for 98943
%./220207/98943 - valid dataset, reduced and included
%./220216/98943 - valid dataset, reduced and included
%./220223/98943 - valid dataset, reduced and included 
%./220306/98943 - the asteroid is g~20 mag, the SNR is very low, data was reduced with Photometry Pipeline (maybe some binning procedure at image level might be required to if we really want to use it)
%./221222/98943 - best dataset we have, reduced and included
%./230107/98943 - the Moon is very close to the objects, the gradients are still in place, the photometric calibration is doubtfull, I suggest to skip it
%./230220/98943 - valid dataset (about 20 -30 min), should be included, data is reduced with Photometry Pipeline
%./230322/98943 -  the SNR is very small (g ~ 20+ mag), data was reduced with Photometry Pipeline (maybe some binning procedure at image level might be required to if we really want to use it)
%./231215/98943 - valid and good dataset, data is reduced and should be included. Minor issues with some of calibrations for some zeropoints
The observations with TCS were made during six observing nights. The data were acquired while the object was at phase angles between 10.6$^\circ$ and 87.5$^\circ$. During each of the nights of December 23, 2022, and December 15, 2023, we continuously imaged the object for more than its rotation period. The observations log is provided in Table~\ref{TCSlog}.

\begin{table}
\centering
\caption{The observations log for the INT telescope. The UT time interval, the ephemeris parameters, and the number of images obtained with B, V, and R filters are shown.}
\label{INTlog}
\begin{tabular}{l c c c c c c c}\hline\hline
UT & $V_{\rm mag}$ & $r$ [au] & $\Delta$ [au] & $\alpha$ [$^\circ$] & $N^B_{\rm img}$ & $N^V_{\rm img}$ &$N^R_{\rm img}$\\
2022-03-03T04:47 -- 06:35 & 18.8 & 0.948 & 0.229 & 94.3 & 13 & 11  & 33\\
\hline
\end{tabular}
\end{table}

During the morning of March 03, 2022, we used the 2.54~m Isaac Newton Telescope (INT) equipped with the Wide Field Camera (WFC) instrument to obtain observations with the $B$ (central wavelength of 430 nm), V (540 nm), and $R$ (640 nm) Johnson/Cousins broadband filters. These observations aimed to obtain the color indexes in these broadband filters. Thus, the observing strategy was to change the filter after each image in the order B - R - V - R - B and so on. This strategy allows a linear interpolation of the magnitude value between the two consecutive observations made in the V to avoid systematic offsets in the colors due to light-curve variation. WFC was mounted at the prime focus of the INT. It was an optical mosaic imaging instrument made of four CCDs, with a resolution of 0.33 arcsec/pixel. For our needs, we only used the central detector, CCD4. The observations log is provided in Table~\ref{INTlog}.

%During the nights of December 15, 16, and 25, 2023, the TTTs telescopes obtained a few images using the $g$, $r$, $i$, and $z_s$ Sloan filters. These could be directly compared with the data obtained with the TCS telescope.

All the spectro-photometric data were reduced using PP.  For astrometric registration, we used the GAIA - DR2 catalog. In the case of TCS, because of its small field, the registration algorithm was applied twice for each image (we noticed that it improved the astrometric accuracy). We discarded those images for which the astrometric registration failed. The PanSTARRS catalog \citep{2012ApJ...750...99T, 2016arXiv161205560C} was used for all photometric calibration. Because of the TCS small field, the accuracy of the photometric calibration depends on the number of stars imaged by each exposure. Thus, for a large number of images, the zero-point error represents the uncertainty of the derived colors.

\begin{table}
\centering
\caption{The log for the spectroscopic observations. The mid-UT time for the observations of  Torifune ($UT_{avg}$ ), the estimated apparent V magnitude, the phase angle ($\alpha$ [$^o$]), the airmass (A.M.), the total number of individual exposures (N$_{exp}$), the exposure time for each image, and the solar analogues are shown.}
\label{SpectraLog}
\begin{tabular}{l l c c c c c c c}\hline\hline
Instrument & UT$_{\rm avg}$       & $V_{\rm mag}$ & $\alpha$ [$^\circ$] &  A.M. & N$_{\rm exp}$ & Exp [s] & Solar analogues\\
GTC        & 2022-12-02T03:45 & 19.2      & 45.7            &  1.16  & 3   &  300     & SA98-978, SA102-1081\\
IRTF       & 2023-02-10T11:59 & 16.3    & 40.0            &  1.5  & 16   &  120    & Hyades 64, HD 75488\\
\hline
\end{tabular}
\end{table}

\subsection{Spectroscopy} \label{subsec:spectro}

The spectral interval 0.5 -- 2.45 $\mu m$ provides key information about the surface composition of the asteroids. It includes the two prominent bands at $\sim$1 and $\sim$2 $\mu m$ characteristic for the olivine-pyroxene mixtures. To obtain high SNR spectra over these wavelengths we used two instruments, GTC for the visible part and IRTF for the near-infrared part. The observations log is shown in Table~\ref{SpectraLog}. Although the two spectra were obtained three months apart, the phase angles were similar (40$^\circ$ compared to 46$^\circ$ ). The visible and the near-infrared parts were merged based on normalization over the common wavelength region (0.82 -- 0.9) $\mu m$, where the curve shapes overlap.

The visible spectrum of Torifune was obtained on the night of December 1, 2022, using the OSIRIS camera-spectrograph\citep{Cepa2000,Cepa2010} mounted at GTC. We used the 1.2" slit with the R300R grism which allows to cover 0.48--0.92~$\mu$m wavelength range. This setup provides a resolution $R$=348 (for a 0.6" slit),  and a dispersion of 7.74 \AA~pixel$^{-1}$. The slit was oriented along the parallactic angle to minimize the effects of atmospheric differential refraction, and the telescope tracking was set at the asteroid's proper motion. Three spectra of 300~s of exposure time each were obtained, with an offset of 10" in the slit direction in between them. To obtain the asteroid reflectance spectrum, we observed two solar analog stars from the Landolt catalog \citep{Landolt1992}, SA98-978 and SA102-1081. The SA98-978 was observed at an airmass  of 1.14, just before the asteroid observations, and SA102-1081 was observed at 1.42 airmass one hour later.  The log of these observations is provided in Table~\ref{SpectraLog}. 

The data reduction was completed using standard procedures. The images were bias subtracted and flat-field corrected. The sky background was subtracted. Then, the one-dimensional spectrum was extracted using a variable aperture defined by the pixels where the intensity reached 10\% of the peak value. Wavelength calibration was carried out using calibration images acquired with the same setup of Xe+Ne+HgAr lamps. This procedure was applied to the spectra of the asteroid and the solar analogs. Then we divided the asteroid spectral curve with one of each solar analogs, and the resulting ratios were averaged to obtain the final reflectance spectrum of Torifune.

The near-infrared spectrum was obtained on the night of February 10, 2023, using the NASA IRTF/SpeX \citep{2003PASP..115..362R}. This is a medium-resolution spectrograph with the CCD camera Teledyne Hawaii-2RG (2048 $\times$ 2048 pixels). Each pixel has 18 $\mu m$, equivalent to a spatial scale of 0.1 arcsec/pixel. We have used this spectrograph in a low-resolution mode -- with the prism diffraction element, covering the (0.8-–2.5) $\mu m$ wavelength interval and a slit of 1.6 $\times$ 15 arcsec.

The spectra of our targets were obtained alternatively on two separate locations on the slit, denoted A and B (called an AB cycle), following the nodding procedure. The exposure time for an individual image of asteroids spectrum was 120 sec. We acquired 16 images (8 AB pairs) for Torifune (Table~\ref{SpectraLog}). The solar analogs Hyades 64 and HD 75488 were observed multiple times throughout the night, as they were positioned near the asteroid in the sky. We selected the images taken with an airmass difference of less than 0.1 compared to the asteroid observations. The calibration images (bias, flat, arc) were obtained at the beginning and end of the observing session. 

To reduce the IRTF spectra, we have used the SPEXTOOL pipeline \citep{2004PASP..116..362C, 2003PASP..115..389V, 2004PASP..116..352V}. After the pre-processing (the bias and flat field corrections), the A and B exposures are subtracted, and the resulting A-B images are added together. This procedure removes most of the sky background. The wavelength calibration is made using the emission lines of an Ar lamp. The possible wavelength shifts (which generate the so-called heartbeats – the reflectance spectral curve shows a positive spike followed or preceded by a negative one because of the wavelength miss-alignment) between the asteroid and the solar analogs are corrected. These are in the order of one pixel.

\section{Results} \label{sec:results}

The above observations allow a detailed ground-based characterization of Torifune's properties before Hayabusa2$\#$ arrival. The extended photometric data set determines the absolute magnitude $H$, the rotation period, the convex shape model and constrains the pole orientation. Then, from the spectro-photometric measurements, we can compute the color indexes and the corresponding object classification, search for a large-scale heterogeneity, and quantify the phase reddening effect.  Finally, the spectral curve indicates the surface composition and the precise spectral type. We can estimate the equivalent diameter by assuming a typical albedo for the assigned taxonomic classification and considering the determined $H$.

\subsection{The absolute magnitude}

The absolute magnitude ($H$) is a fundamental astronomical parameter that links the asteroid's size and its surface albedo. The $H$ is  defined as the mean apparent magnitude in the Johnson V band over a full rotation cycle when observed at a distance 1 au from both the Sun and the Earth and at a phase angle of 0$^\circ$. In practice, observing asteroids at 0$^\circ$ phase angle ($\alpha$) is very difficult. Moreover, the curve of brightness as a function of phase angle $\alpha$ is non-linear for $\alpha < 7^\circ$ \citep[e.g.,][]{2010Icar..209..542M}. Thus, several empirical models were derived to accomplish this task. Among the most popular models are the $\{H$, $G\}$ and  $\{H$, $G_1$, $G_2\}$ \citep{2016P&SS..123..117P}. Below, we provide a short summary of the equations used to apply these models to our data following \citet{2010Icar..209..542M}. 

\begin{equation}
    V_{\rm red}(\alpha) = V - 5 \cdot \log(r \cdot \Delta)
    \label{Hred}
\end{equation}

\begin{equation}
    H = V_{\rm red}(\alpha) + 2.5 \cdot [\log_{10}(1-G) \cdot (\phi_{1}(\alpha)) + G(\phi_{2}(\alpha))]
    \label{Habs}
\end{equation}

\begin{equation}
    \phi_{i}(\alpha) = \exp(-A_i \cdot (\tan^{B_i}(\alpha/{2}))
    \label{Phis}
\end{equation}

The reduced magnitude (Eq.~\ref{Hred}) can be expressed as a function of the apparent magnitude ($V$), the heliocentric distance ($r$), and the geocentric distance $\Delta$ (where both $r$ and $\Delta$ are expressed in au). Following this equation, the relation between the reduced magnitude and the absolute magnitude can be written (Eq.~\ref{Habs}) by taking into account the slope parameter ($G$) and $\alpha$. The two functions of $\phi_{1}(\alpha)$ and  $\phi_{2}(\alpha)$ can be expressed by the Eq.~\ref{Phis} and the coefficients $A_1$ = 3.33, $A_2$ = 1.87, $B_1$ = 0.63, $B_2$ = 1.22 \citep{2010Icar..209..542M}. 

\begin{equation}
H = V_{\rm red}(\alpha) + 2.5 \log_{10} \left[ G_1 \Phi_1 (\alpha) + G_2 \Phi_2 (\alpha) + (1 - G_1 - G_2) \Phi_3 (\alpha) \right],
\label{hg1g2}
\end{equation}

\citet{2010Icar..209..542M} noted that $\{H$, $G\}$ model  does  a reasonably fit for most of the asteroids in the phase angles interval from $\sim10^\circ$ to $\sim60^\circ$. Although the $\{H$, $G\}$ model has the advantage of simplicity, achieving a better fit to high-quality photometric phase data requires introducing an additional parameter to the photometric phase function. \citet{2010Icar..209..542M} proposed a three-parameter $\{ H$, $G_1$, $G_2\}$ magnitude phase function for asteroids (Eq.~\ref{hg1g2}) which depends  on the three phase functions, $\Phi_1$, $\Phi_2$, and $\Phi_3$.

\begin{figure}
	\includegraphics[width=12cm,keepaspectratio]{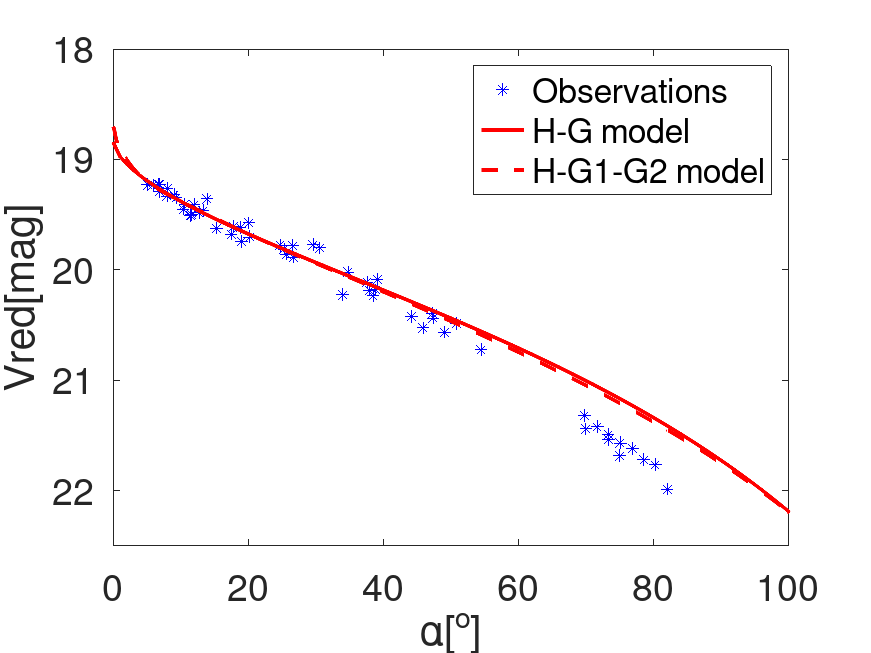}
    \caption{The reduced magnitude  $V_{\rm red}$ as a function of phase angle $\alpha$. Each of the blue points represents the average of the observations made with an instrument during a night. The log of these observations is shown in Tabel~\ref{photomlogtable}. The fitted red curve corresponds to  $\{H$, $G\}$  model where $H = 18.846$ and $G = 0.294$, while the red dashed curve represents the $\{H$, $G_1$, $G_2\}$ model where $H = 18.706$, $G_1 = 0.259$, and $G_2 = 0.372$.}
    \label{fig:HG}
\end{figure}

Our observations were performed with multiple instruments and filters. Most of the data points were observed with TTT2 telescope equipped with QHY411 camera and the Sloan $g$ filter. As the first step for data preparation we computed the magnitude offsets between the data acquired with all other instruments and the TTT2/QHY411/$g$-filter. This was possible because some observations were performed simultaneously with various instruments. The errors introduced by this transformation are negligible gracefully to the large number of points used to compute the offsets.  

The conversion between the $Luminance$ filter and the $g$ filter is performed using Eqs.~\ref{LuminaceI} and \ref{LuminaceQ}. This calculation relies on the $(g-r)$ color (reported as 0.663 mag in Section 3.3) and assumes that the asteroid's surface is homogeneous on a large scale, meaning its colors do not vary during rotation.

\begin{equation}
    (V-g) = 0.005 - 0.536\cdot(g-r) + 0.011 \cdot(g-r)^2 
    \label{vmg}
\end{equation}

The second step in the data preparation involved converting the asteroid's apparent $g$ magnitudes to the $V$ band, as required for the definition of the $H$ magnitude. Since we used the Pan-STARRS1 photometric system to reduce our data, we applied the conversions outlined by \citet{2012ApJ...750...99T}. The value of $(V-g)$ can be expressed as a second-order polynomial, according to Table 6 from \cite{2012ApJ...750...99T}. Using the $(g-r)$ value of 0.663 mag found in Section 3.3, we calculated $(V-g) = -0.345 \pm 0.03$. This offset was applied to convert all photometric values to the $V$ magnitude, under the assumption that Torifune's colors do not vary during the rotation, as also indicated in Section 3.3.

%Because the asteroid shows a surface homogeneity (at large scale, as it will be shown in Section 3.1), this transformation is also a magnitude offset -- the $(g-V)$ color.  This value can be computed from the asteroid spectrum with the help of the synthetic photometric technique \citep[e.g.][]{2018A&A...617A..12P} -- Section 2.1, although it refers to other spectral interval, it is valid in general.  One issue  for applying this technique is that available spectrum starts at 0.5~$\mu m$, while the $g$ band starts at $\sim0.4~\mu$m. This was solved by linearly extrapolating the spectrum (we fitted the spectrum with line over 0.5 - 0.6 $\mu m$ and extrapolated it down to 0.4 $\mu$m). This methods introduces an error less than $0.02$ magnitudes (computed as the standard deviation of the results obtained trough various extrapolation approaches). In addition to this synthetic photometric value of $0.11\pm0.02$ mag (computed with the reflectance spectrum), we must take into account $(g-V)_{Sun}$ = 0.42 mag \citep{2018ApJS..236...47W}. Thus we determined $(g-V) = 0.53$ mag for Torifune. This offset was used to convert all the photometric values to the V magnitude. 

\begin{equation}
    \cos \Theta = \sin \delta \sin \delta_0 + \cos \delta \cos \delta_0 \cos (\alpha - \alpha_0)
    \label{aspectangle}
\end{equation}

Furthermore, our observations were conducted from late 2022 to early 2024. As a result, it is important to account for the changing geometry of the observations \citep{2024A&A...687A..38C}. This can be achieved by considering the aspect angle ($\Theta$). The average values of $\Theta$ for each observing night are reported in Table~\ref{photomlogtable2}. These values were computed using Eq.\ref{aspectangle}, where $(\alpha,\delta)$ represent the equatorial coordinates of the asteroid at the time of observation, and $(\alpha_0,~\delta_0)$ denote the equatorial coordinates of Torifune's spin axis, as computed in Section 3.2. To calculate the $H$ values, we only used observations where $93^\circ \geq \Theta \geq 79^\circ$. This interval is comparable to the uncertainty in $\Theta$ (caused by the propagation of the pole's uncertainity) and encompasses a significant portion of our observations.

The computations were made using the Pyedra\footnote{\url{https://pyedra.readthedocs.io/}} \citep{2022A&C....3800533C} Python package.  The result is  $H = 18.846 \pm 0.072$ mag, and  $G = 0.283 \pm 0.025$. 
The {$H$, $G_1$, $G_2$} model was also applied on this sub-sample using the web-interface \footnote{\url{ https://psr.it.helsinki.fi/HG1G2/}} provided by \citet{2016P&SS..123..117P}. The reported values are $H = 18.706$, $G_1 = 0.259$, and $G_2 = 0.372$. The $G_1$ and $G_2$ values for which the target is best explained are typicall for an S/M type. The simulated confidence interval for $H$, according to \citet{2016P&SS..123..117P} is $H \in (18.32, 19.09) $. The Fig.~\ref{fig:HG} shows the matching between the observations and the two models, $\{H$, $G\}$ and $\{H$, $G_1$, $G_2\}$. The measurement errors can be statistically estimated as the dispersion of points around the general trend. They are about $\sim$0.06 mag. The difference between the two models can be attributed to the lack of observational data near a $0^\circ$ phase angle. In conclusion, we will take the average value of the two models and use the difference between them as the uncertainty, resulting in $H = 18.78 \pm 0.14$.

\begin{equation}
\Delta_{mag}^\Theta = 2.5 \log_{10} \left[ 1 - (1 - R)|\cos \Theta| \right].
\label{deltamag}
\end{equation}

There is a miss-match between the models and the observations made at high phase angles ($\alpha~\geq~69^\circ$). All these data points were determined at an aspect angle $\Theta \approx 50 ^ \circ$.  This difference, seen in Fig~\ref{fig:HG} can be explained by Eq.~\ref{deltamag} \citep{2024A&A...687A..38C} where $R$ is polar-to-equatorial oblateness ($0~<~ R~\leq~1$).

%Because all the observations were made with various telescopes and cameras (table~\ref{photomlogtable}) and the data has been reduced with multiple methods we can exclude an uncounted system error. All the observational dataset can be matched with the {$H = 18.60$, $G_1 = 0.961$, $G_2 = 0.016$}, but the values of $G_1$ and $G_2$ are not realistic  for an S-type asteroid.  Most of the observations at low phase angles were obtained in Novemeber and Decemeber 2023, while those at large phase angles were observed in March 2023. The different observing geometries may account for the observed miss-match.

Our determined $H$ value is comparable  with the one ($H=18.74$) currently listed by the JPL Horizon system (the web page does not report an associated uncertainty). However, \citet{2024A&A...688L...7F} reported an absolute magnitude of $H = 18.94 \pm 0.05$ based on six observations made in 2005, 2022, and 2024, along with apparent magnitude values from the astrometric reports in the MPC database. These authors did not account for the varying observing geometry, which can affect the calculation for oblate bodies.

%. The differences can be explained by the amount of points used to fit the {$H$, $G$} or {$H$, $G1$, $G2$} models and by the fitting accuracy.

%Since our photometric observations were obtained at $5.1^\circ\leq\alpha\leq82.1^\circ$ we applied the {$H$, $G$} model (Fig.~\ref{fig:HG}).  To increase the accuracy, we used only the data acquired for $\alpha\leq40^\circ$.  Because most of our data were acquired with the Sloan filters, we did all our computations for the $g$ filter. Thus, we found $H_g = 19.147 \pm 0.010$ mag, and  $G = 0.273 \pm 0.004$.

%The last step is to find the offset between $H_g$ and $H$ (which is referenced with respect to the Johnson V filter). This can be written as $H$ = $H_g$ - $(g-V)$. Since we know the asteroid reflectance spectrum, we can derive and compute the $(g-V)$ using the filters profile and the corresponding solar color, $(g-V)_{Sun}$ = 0.33 mag \citep{2018ApJS..236...47W}. Thus, after performing synthetic photometry on the spectrum and by adding the solar colour, we found $(g-V) = 0.45$ mag for 2001 CC$_{21}$.
%Consequently we found $H ~ = ~18.693 ~\pm ~0.010$.  

%%%%%%%%% SHAPE Model Section
\subsection{The rotation period, the convex shape model, and the pole orientation}

The high SNR lightcurves obtained through our observing survey allow us to determine the rotation period (P),  the pole orientation -- its ecliptic longitude ($\lambda$), ecliptic latitude ($\beta$), and obliquity ($\epsilon$), and to estimate the morphological model. These computations followed the same approach as \citet{2024MNRAS.527.6814R}. We used the code publicly available in the Database of Asteroid Models from Inversion Techniques (DAMIT\footnote{\url{https://astro.troja.mff.cuni.cz/projects/damit/}}; \cite{2010A&A...513A..46D}).

The sidereal rotation period (P) is the time the asteroid takes to complete a single rotation over its axis, with respect to the the background stars.  There is a large number of algorithms for the rotation period determination \citep[][ and references there in]{2018ApJS..236...16V}. These can use Fourier analysis or the optimization of a cost function  across candidate periods (i.e phase folding, least-squares, Bayesian approaches). We applied a search for the rotation period using the $\chi^2$ cost function.

\begin{figure}
    \centering
    \includegraphics[width=12cm,keepaspectratio]{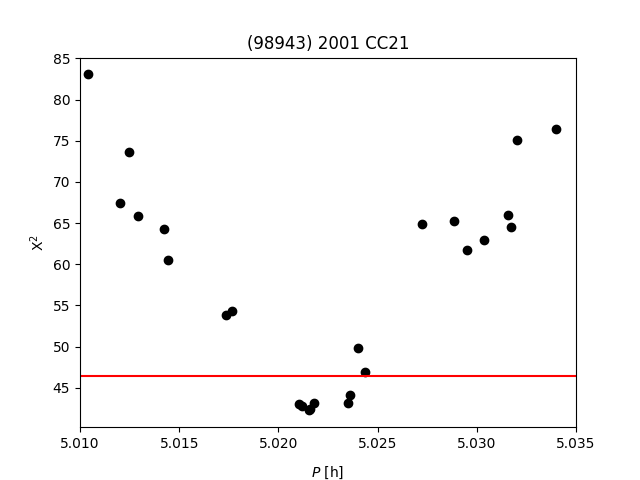}
    \caption{Period search tool output plot for (98943) Torifune. This search was performed in an interval from 5.010 h to 5.035 h, with a coefficient $p$ (period step) of 0.2. Each obtained period is represented as a black dot, with the red line representing a 10\% threshold from the lowest $\chi^{2}$ obtained. The presence of 5 values under this threshold is an indicator that more data will refine the period value.}
    \label{fig:period_plot}
\end{figure}

We employed a set of 21 lightcurves with a temporal span of 100 days, from 26 November 2022 to 6 March 2023.  All these where light-time corrected using the JPL Horizons ephmerides. As an initial guess for the period search, we took the values $P~=~5.022 \pm 0.001$ h \citep{FatkaPGC} and $P~=~5.0159 \pm 0.0006$ h \citep{2023MPBu...50..217W}.  A similar value ($P~=~5.02124\pm0.00001$ h) was reported by \citet{2024A&A...688L...7F}. We performed several searches around these values, finding a minimum close to $P=5.021521$ h in most cases, as shown in Figure \ref{fig:period_plot}. This value was adopted as the initial period of the code. 

%This asteroid's period was previously in (Pravec 2002web\footnote{\url{http://www.asu.cas.cz/~asteroid/2001cc21_2002.png}} and 2022web) obtaining the following periods respectively: $5.017 \pm 0.001$ h and $5.0247 \pm 0.0001$ h; and in \cite{2023MPBu...50..217W} with a period of $5.0159 \pm 0.0006$ h.We took those values as a starting point to the period tool provided with the code, 
\begin{figure*}
    \centering
    \includegraphics[width=\textwidth]{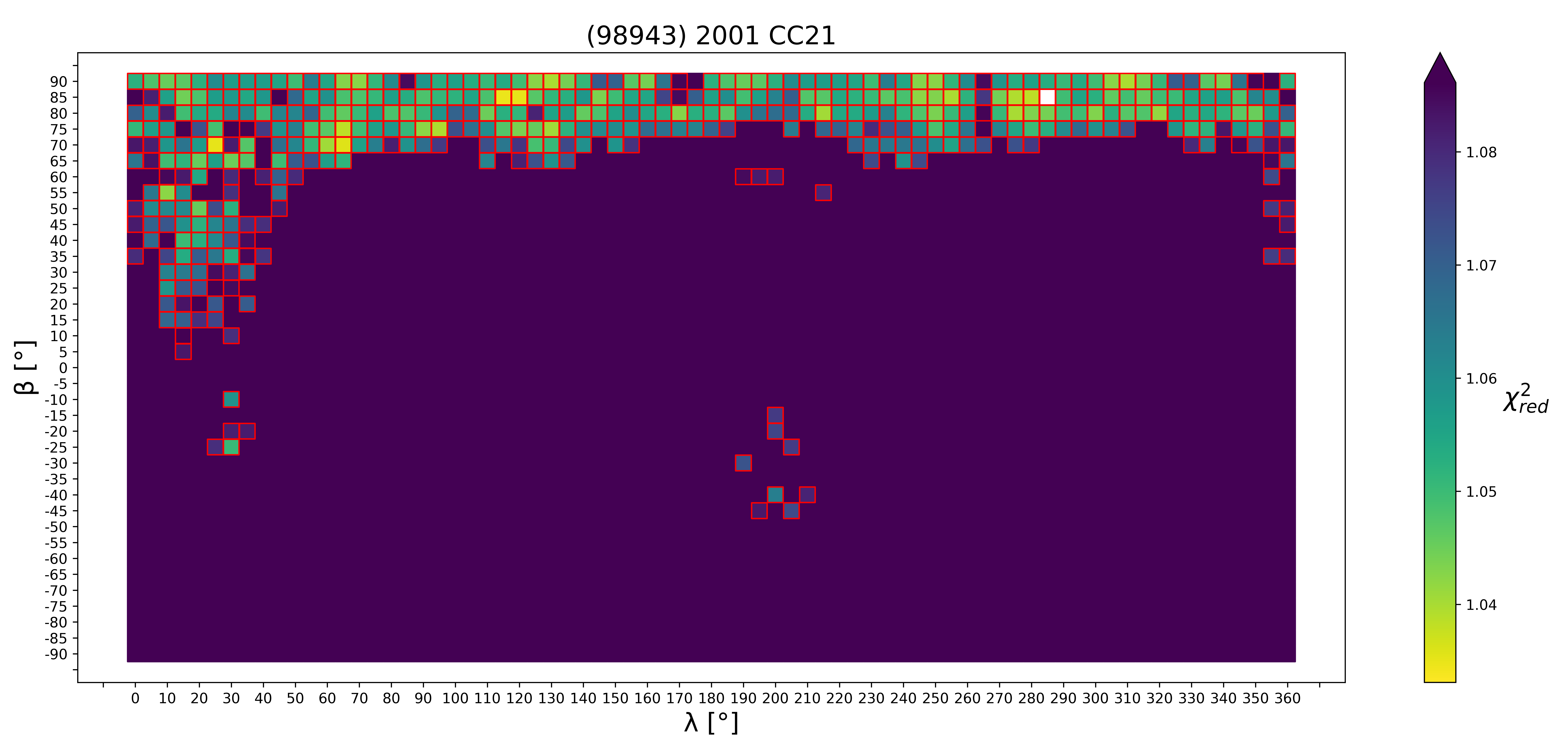}
    \caption{Pole solutions statistical quality for (98943) Torifune. The values of $\lambda$ and $\beta$ are shaded by its $\chi_\mathrm{red}^{2}$ value, while the best solution, ($\lambda=280 ^{\circ}, \beta=85 ^{\circ}$ with $\chi_\mathrm{red}^{2}=1.04$) is plotted as a white square. The red border represents the solutions that are within a 3$\sigma$ level of uncertainty from the best solution of 4.88 \%.}
    \label{fig:pole_plot}
\end{figure*}

In the second step, we run the medium resolution search across the entire sphere ($0^{\circ} < \lambda \leq 360^{\circ}, -90^{\circ} \leq \beta \leq 90^{\circ}$) in 5$^{\circ}$ steps, obtaining a solution of $P=5.021425$ h, $\lambda = 280^{\circ}$, $\beta = 85^{\circ}$ with a $\chi_\mathrm{red}^{2}=1.04$. In Figure \ref{fig:pole_plot}, we show a representation of the values obtained for all the solutions shaded by their $\chi_\mathrm{red}^{2}$ value.

\begin{figure*}
    \centering
    \includegraphics[width=7cm]{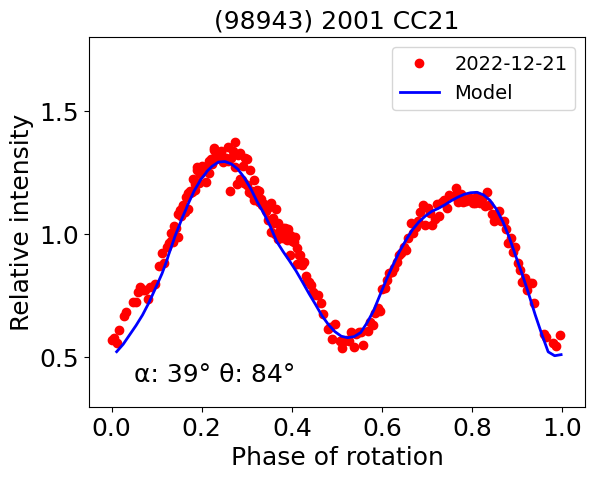}
    \includegraphics[width=7cm]{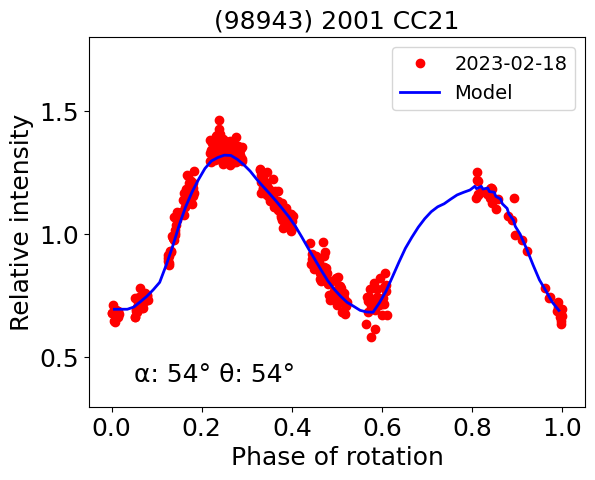}
    \includegraphics[width=7cm]{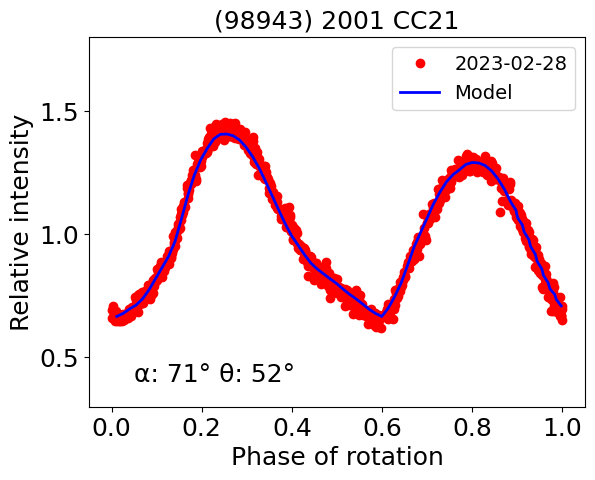}
    \includegraphics[width=7cm]{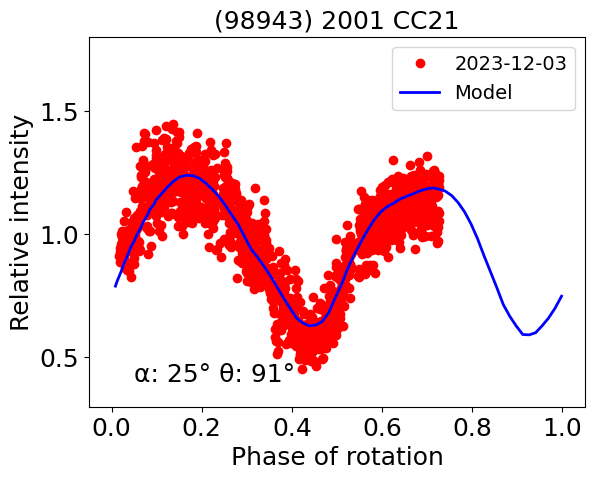}
    \includegraphics[width=7cm]{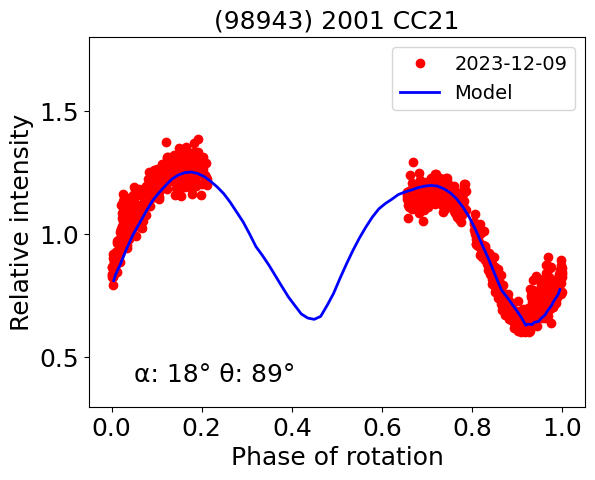}
    \includegraphics[width=7cm]{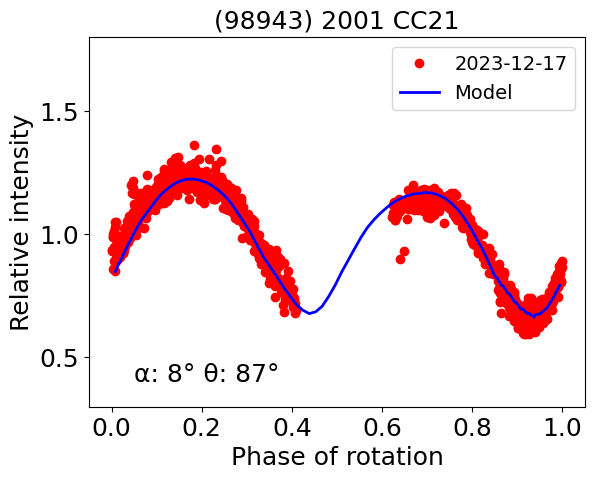}
    \caption{Graphical representation of the fit between four light curves and the best-obtained model for Torifune. The observed data is plotted as red dots, while the shape model is plotted as a solid blue line for each observation. The geometry is described by its solar phase angle $\alpha$ (defined as the angle between the Sun, the asteroid and the observer) and its aspect angle $\theta$ (defined as the angle between the rotation axis and the asteroid - observer direction). The reference Julian Day is 2459909.000000.}
    \label{fig:lc_fit_sel}
\end{figure*}

\begin{figure}
    \centering
    \includegraphics[width=12cm,keepaspectratio]{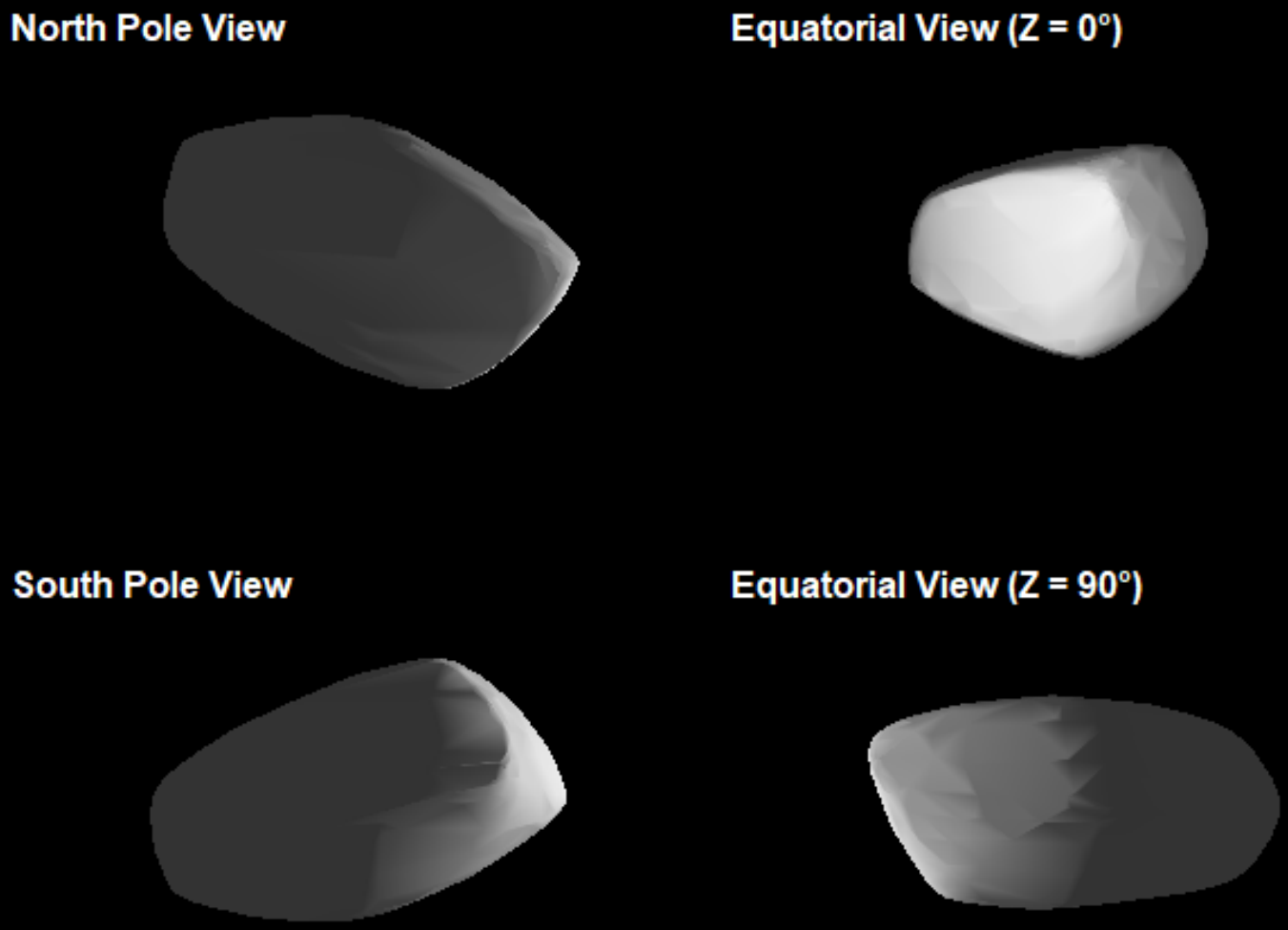}
    \caption{Obtained shape model of (98943) Torifune. Left top: North Pole View (Y axis = 0$^{\circ}$). Left bottom: South Pole View (Y axis = 180$^{\circ}$). Right top: Equatorial View with Z axis rotated 0$^{\circ}$. Right bottom: Equatorial View with Z axis rotated 90$^{\circ}$.}
    \label{fig:shape_model}
\end{figure}

In the last step, we made a fine search around the previously obtained values. We probed in an interval of 30$^{\circ} \times$30$^{\circ}$ square in 2$^{\circ}$ steps. We obtained $P=5.021512$ h, $\lambda = 279^{\circ}$, $\beta = 88^{\circ}$, $\epsilon = 4^{\circ}$ with a $\chi_\mathrm{red}^{2}=1.03$. The value of $\epsilon$ implies a prograde rotation. Figures \ref{fig:lc_fit_sel} and \ref{fig:lc_fit_rest} show a graphical representation between the fit of the lightcurve data and the shape model, while Figure \ref{fig:shape_model} shows the obtained shape model of the asteroid. The ecliptic latitude of the best fitting model, with a value near the 90$^{\circ}$, makes it possible for another pole solution to be valid, with a change in the ecliptic longitude of 180$^{\circ}$, that is, $\lambda \simeq 110^{\circ}$.

\begin{figure}
    \centering
	\includegraphics[width=9cm]{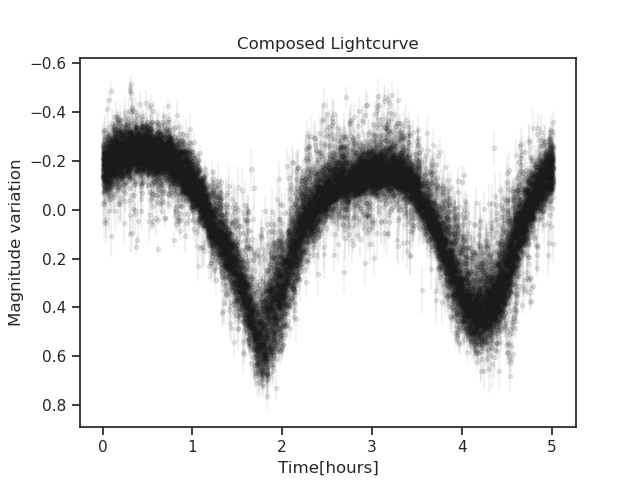}
    \caption{The folded lightcurve for a period of $P= 5.021516 \pm 0.000106 $ h. We selected the data with phase angles between 4$^\circ$ and  23~$^\circ$.  The reference Julian day is 2460286.642715431.}
    \label{fig:FLC}
\end{figure}

For the uncertainties estimation, we made 100 subsets from the main data set ($\sim 7000$ measures) by randomly removing 25\% of the data. For each of them, we run a fine search around the best medium search solution ($P=5.021425$ h, $\lambda = 280^{\circ}$, $\beta = 85^{\circ}$), obtaining thus 100 solutions. We obtained the following results, $P= 5.021516 \pm 0.000106 $ h (the folded lightcurve is represented in Fig.~\ref{fig:FLC}), $\lambda = 301^{\circ} \pm 35^{\circ}$, $\beta = 89^{+1}_{-6}~^{\circ}$ and $\epsilon = 5^{\circ} \pm 3^{\circ}$. 

%There are also light curves available on the the Asteroid Lightcurve Data Exchange Format (ALCDEF) database with a temporal span of 10 days between 22 January 2023 and 1 February 2023, which is within our temporal span. We decided not to add them to our model since the solution were not improved.

\begin{table}
\centering
\caption{The median $(g-r)$, $(r-i)$, $(i-z_s)$, and $(r-z_s)$ colors of the Torifune are provided for each observing session. The phase angle ($\alpha$) column is repeated from Table~\ref{TCSlog} to allow an assessment of the phase reddening effects. }
\label{TCSColors}
\begin{tabular}{l c c c c c c c c c}\hline\hline
Obs. date  & $\alpha$ [$^\circ$] & $(g-r)$ & $(g-r)_{err}$ & $(r-i)$ & $(r-i)_{err}$  & $(i-zs) $ & $(i-zs)_{err} $  & $(r-z_s) $ & $(r-z_s)_{err} $\\ \hline
2022-02-08 & 71.8            & 0.662 & 0.050 &       &       &        &       & 0.083 & 0.047\\
2022-02-17 & 80.3            & 0.638 & 0.051 &       &       &        &       & 0.128 & 0.055\\
2022-02-24 & 87.5            & 0.661 & 0.093 &       &       &        &       & 0.141 & 0.099\\
2022-12-23 & 37.5            & 0.686 & 0.059 & 0.182 & 0.054 & -0.044 & 0.060 & 0.135 & 0.058\\
2023-02-20 & 57.7            & 0.690 & 0.041 & 0.186 & 0.040 & -0.097 & 0.045 & 0.085 & 0.044\\
2023-12-15 & 10.6            & 0.638 & 0.049 & 0.164 & 0.047 & -0.041 & 0.048 & 0.123 & 0.049\\
\hline
\end{tabular}
\end{table}
%%%%%%%%%%%%%%%%%%%%%%%%%%%%%%%%%%%%%%%%%%%%%%%%%%

\subsection{Colors and surface homogeneity}

\begin{figure}[ht!]
	\includegraphics[width=9cm,keepaspectratio]{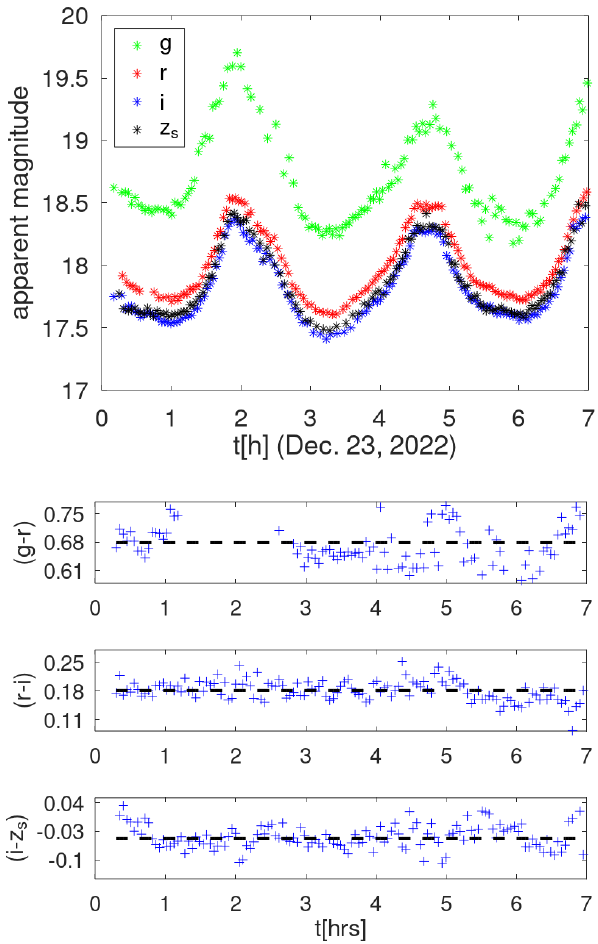}
	\includegraphics[width=9cm,keepaspectratio]{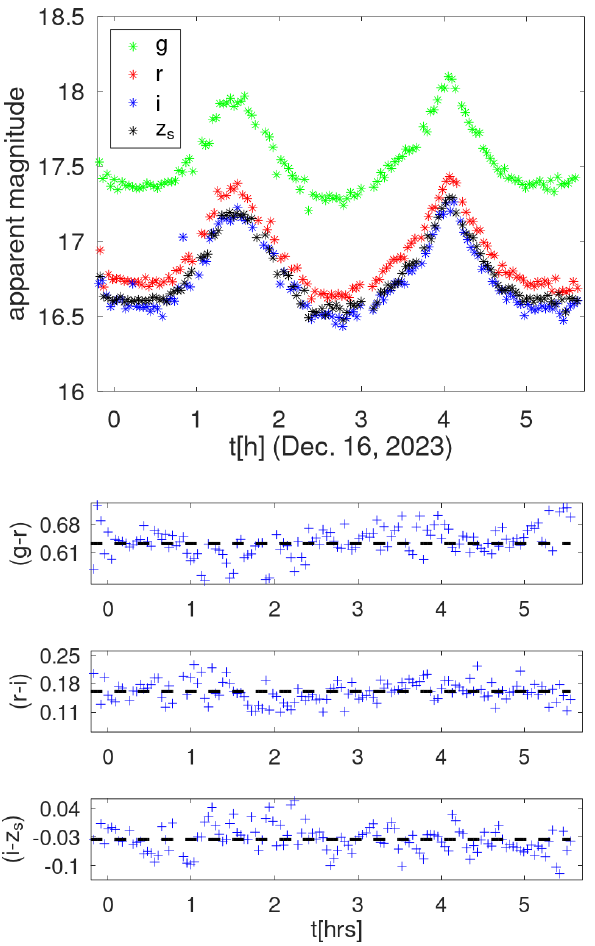}
    \caption{The simultaneous lightcurves obtained with g, r, i, and z$_s$ filters. The observations were performed with TCS/MuSCAT2 instrument. The data corresponding to $(g-r)$ for the night of Dec. 23, 2022, has a low SNR due to the magnitude limit of the instrument. }
    \label{fig:TCS}
\end{figure}

The simultaneous images obtained with different broadband filters acquired with TCS/MuSCAT2 instrument provide the best means to detect a possible large surface heterogeneity. During two nights (December 23, 2022, and December 16, 2023), one year apart, we observed Torifune at TCS telescope for a time longer than one rotation period cycle -- $P=5.021512$ h. The lightcurves are shown in Fig~\ref{fig:TCS}. With these data, we can trace the color curve with respect to time (and consequently with the asteroid observed facet at a time). These plots depicted in Fig~\ref{fig:TCS} show no variation of color -- or a consistent trend (which can be filtered out with a low pass filter), with the time (the curves are flat within the uncertainties). The photometric error bars are similar with the points dispersion \citep{2014sdmm.book.....I}. The noisiest curve corresponds to $(g-r)$ observed on December 23, 2022, and it is due to Torifune's faint  apparent magnitude. The corresponding points were discarded due to their dispersion (which is much larger compared to the scale of the figures). For comparison, the expected colors for an L-type(retrieved from four objects using synthetic photometry) are $(g-r)~=~0.725\pm0.016$ mag, $(r-i)~=~0.215\pm0.014$ mag,	$(i-z_s)~=~0.031\pm 0.030$ mag.

Using each set (the data obtained during an observing night) of  observations made with TCS, we computed the median value for $(g-r)$, $(r-i)$, and $(i-z_s)$ colors. These are reported in Table~\ref{TCSColors}.  At the beginning of 2022, the $i$ channel was not available due to technical issues, thus the $(r-i)$ can't be computed for those dates. The best estimate for the photometric errors is the uncertainty of the photometric calibration constant (zero point) of each image. The error for the instrumental magnitude (the uncalibrated value computed from the aperture photometry method) is negligible because of a large number of images.

To conduct the classification using the colors, we computed the average values and their standard deviation for the overall observation nights. These are $(g-r)~=~0.663\pm 0.022$, $(r-i)~=~0.177\pm 0.012$, $(i-z_s)~=~-0.061\pm0.032$,	and $(r-z_s)~=~0.116\pm0.025$  magnitudes. We notice that for $(g-r)$ color, there is a slight variation (but smaller than the uncertainties) with the phase angle, but the trend is not consistent. Thus, with our data in the visible region, we cannot find a significant phase reddening effect.

We applied the K-Nearest Neighbours (KNN) and the Random Forest (RF) algorithms to complete the classification task. For these, we used the {\tt Python} package {\tt SCIKIT-LEARN}. The KNN algorithm classifies an object based on the class of its neighbours in the color-color diagram, while the RF algorithm assigns the final classification to an object using decision-tree structures. Both algorithms require a training set consisting of asteroids for which we know both the photometric colors and the accurate classification.  To generate it, we searched for the available asteroid classification of all the objects observed with TCS previously. We took the references for the training set from the classifications provided by the SMASS-MIT-Hawaii Near-Earth Object Spectroscopic Survey (MITHNEOS MIT-Hawaii Near-Earth Object Spectroscopic Survey) program \citep{2019Icar..324...41B} and for the Modeling for Asteroids (M4AST) database \citep{2012A&A...544A.130P}. We retrieved spectral classifications for 84 of the NEAs observed as well by our TCS/MuSCAT2 programme \citep{2021EPSC...15..820P}. To increase the training sample, we computed the synthetic colors using the visible spectra published by \citet{2019A&A...627A.124P} and \citet{2018P&SS..157...82P}. The final training sample consisted of 154 asteroids classified as 5 A-types, 8 V-types, 34 Q-types,  48 S-complex (the Sa, Sr, Sv, Sq, and S are grouped together), 7 B-types, 15 C-complex (C, Cb, Cg, Ch, Cgh are grouped together), 9 D-types, and 28 X-complex (X, Xe, Xc, Xk, and Xn are grouped together).

To account for the color uncertainties, we applied a Monte Carlo approach. We started with each color value and its associated uncertainty. We generated three normal distributions (one for each color) of 10\,000 fictitious color values. Then, for each of these cases, we classified the object. Finally, the assigned taxonomy was the one with the highest frequency.

\begin{figure}[ht!]
	\includegraphics[width=18cm,keepaspectratio]{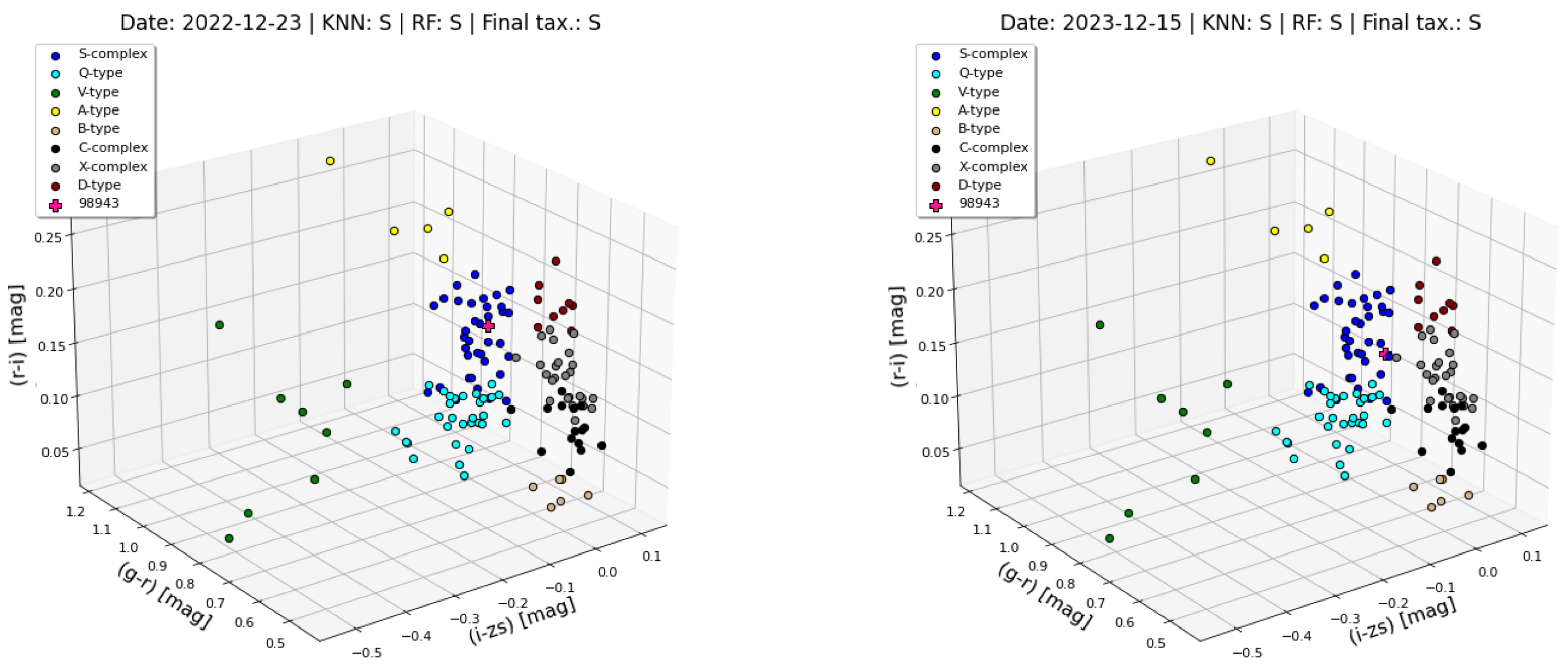}
    \caption{The  $(g-r)$, $(r-i)$, $(i-z_s)$ color-color diagram of the 155 objects with known spectral classification, used as training data. The values corresponding to Torifune are shown in magenta. The title shows the classification according to the K nearest neighbors (KNN) and Random Forest (RF) algorithm. The results corresponding to two observing nights are shown to allow the comparison of Torifune position.}
    \label{fig:TCStax}
\end{figure}

Thus, based on the average color values of $(g-r)~=~0.663\pm 0.022$ mag, $(r-i)~=~0.177\pm 0.012$ mag, $(i-z_s)~=~-0.061\pm0.032$ mag, both algorithms classify this object as an S-complex member with 100\% probability (this probability indicates that all the clones generated within the Monte Carlo approach were classified as S-complex objects). We note that colors do not provide the means to assign undoubtedly one of the sub-types of S-complex (Sa, Sr, Sv, Sq, or S).

The few observations made with INT allowed us to determine the $(B-V)$ = 1.142$\pm$0.09 mag, and $(V-R)$ = 0.461$\pm$0.05 mag colors. These observations were made in the morning of March 3, 2022 when the object was at 94.3$^\circ$ phase angle. The $(B-V)$ value is slightly redder compared to  the values reported by \citet{2024A&A...688L...7F}  who found   $(B-V)$ = $0.95\pm0.06$ and $(B-V)$ = $0.85\pm0.04$ at phase angles around $\sim 47 ^\circ$. The difference may represent a rough estimate of the phase reddening effect for this asteroid. 
 
%%%%%%%%%%%%%%%%%%%%%%%%%%%%%%%%%%%%%%%%%%%%%%%%%%

\begin{figure}[ht!]
	\includegraphics[width=6.01cm,keepaspectratio]{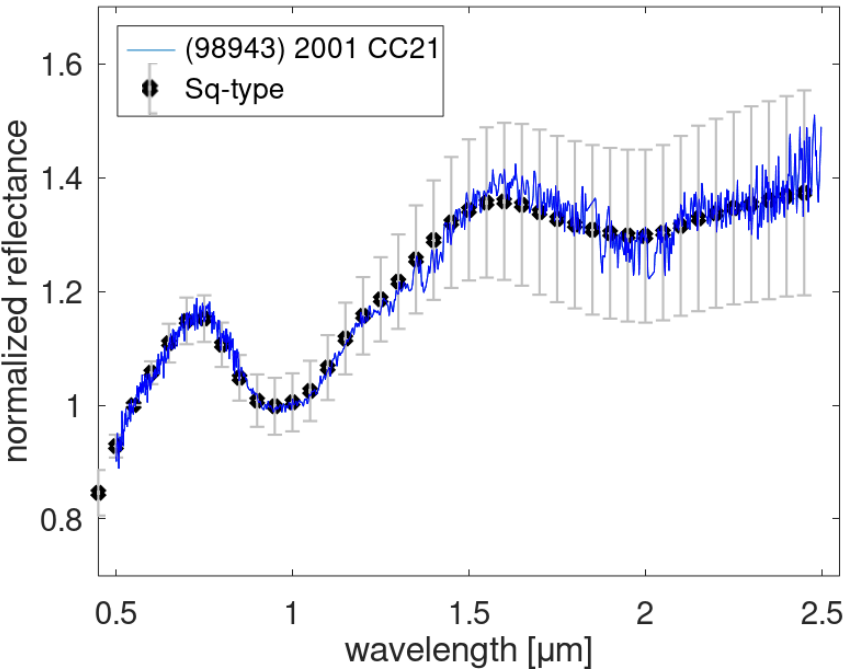}
	\includegraphics[width=6.18cm,keepaspectratio]{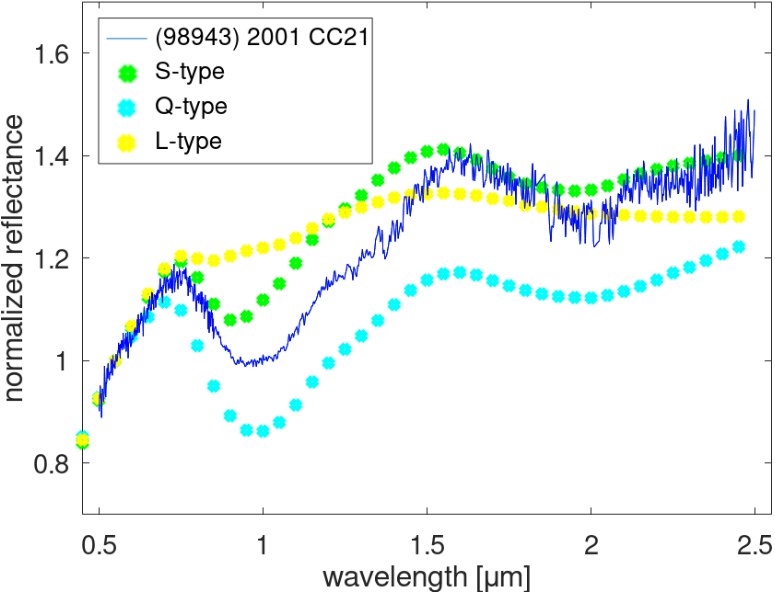}
 	\includegraphics[width=5.95cm,keepaspectratio]{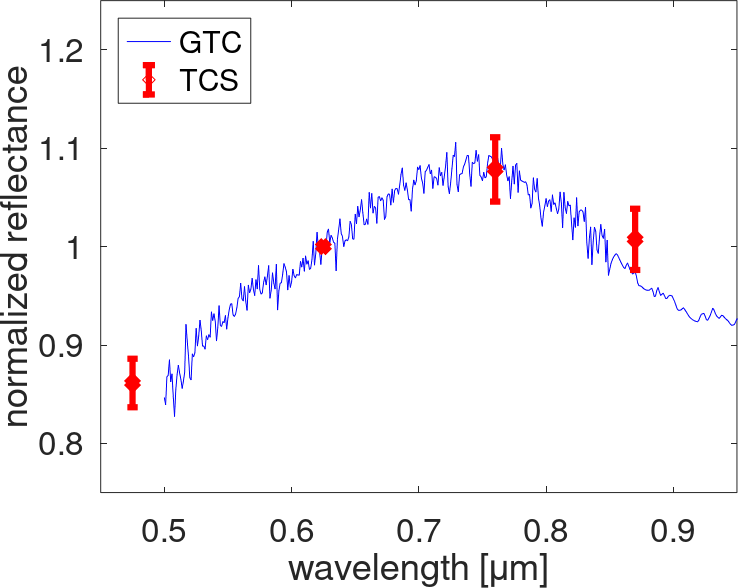}
    \caption{ The comparison between the spectrum of Torifune and the Sq-type (left) and the S-, Q- and L- types (center) as they were defined by Bus-DeMeo taxonomy \citep{2009Icar..202..160D}. (Right) Comparison between the visible spectrum of Torifune obtained with GTC and the spectro-photometric data obtained with TCS. All spectra are normalized at 0.55 $\mu m$.}
    \label{fig:VNIRspecTax}
\end{figure}

\subsection{Spectroscopy and surface composition}

The visible to near-infrared spectral curve of Torifune is shown in Fig.~\ref{fig:VNIRspecTax}. The taxonomic classification was made using the M4AST \citep{2012A&A...544A.130P} and the SMASS - MIT\footnote{\url{http://smass.mit.edu/busdemeoclass.html}} web interfaces. Both tools classify it as an Sq type in the Bus-DeMeo taxonomic system \citep{2009Icar..202..160D} (Fig.~\ref{fig:VNIRspecTax} left).  The difference is a slight shift of the second band minimum, which is attributed to subtle differences in composition.  In Fig.~\ref{fig:VNIRspecTax} (center) we show a comparison with the L- S- and Q- types which were proposed by the previous publications. The region around 1 $\mu m$ band outlines the miss-match of these classes with the spectrum of Torifune. 

NIR spectra of Torifune  were published by \citet{2023MNRAS.525L..17G, 2009Icar..202..160D} and \citet{2005MNRAS.359.1575L}.  A comparison between these spectra is made by \citet{2023MNRAS.525L..17G}. These spectra are similar within the errors, with a very slight difference related to the spectral slope (which can be attributed either to phase angle effects, or to solar analogue used for the data reduction). We compared our spectrum with the one reported by \citet{2023MNRAS.525L..17G}. These two curves are almost identical (Fig.~\ref{fig:98943oursvsGeem}). There is a slight difference in the spectral slope. The 1 $\mu m$ shows a consistent bowl shape four our result, while  the spectrum \citet{2023MNRAS.525L..17G} shows a spike around 0.9 $\mu m$ which might be an artefact.

A cross-validation between the visible spectrum obtained with GTC and the observations obtained with TCS can be made by converting the colors to reflectances. This method requires the colors of the Sun, for which we took the following values, $(g - r)_{\rm Sun}$ = 0.50 mag, ($r - i)_{\rm Sun}$ = 0.10 mag, ($i - z_s)_{rm Sun}$ = 0.03 mag. These values were derived by considering the profiles of the filters available for the MuSCAT2 instrument and are consistent with those provided by \citet{2006MNRAS.367..449H}. The matching (Fig.~\ref{fig:VNIRspecTax} -- right) between the data obtained with two different observing methods provides reliability to our results.

\begin{figure}[ht!]
	\includegraphics[width=9cm,keepaspectratio]{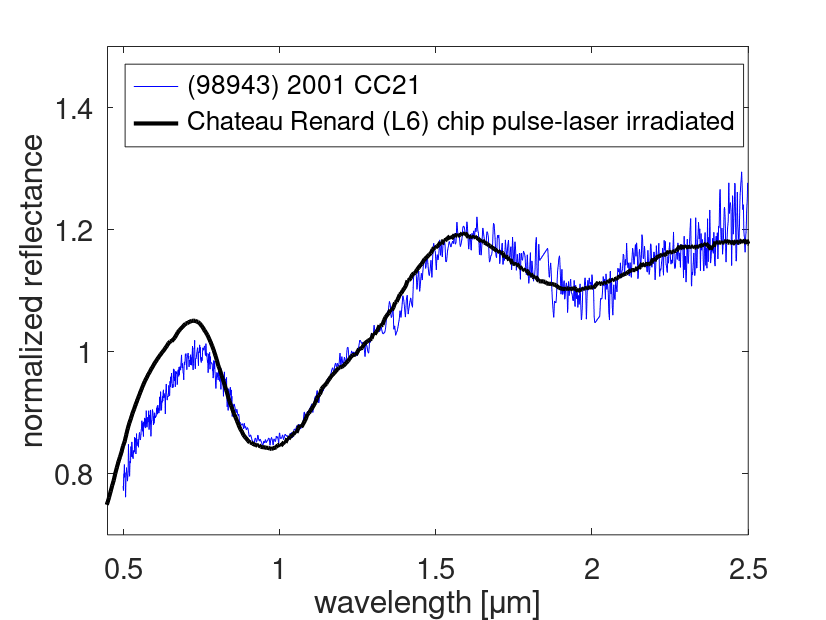}
	\includegraphics[width=9cm,keepaspectratio]{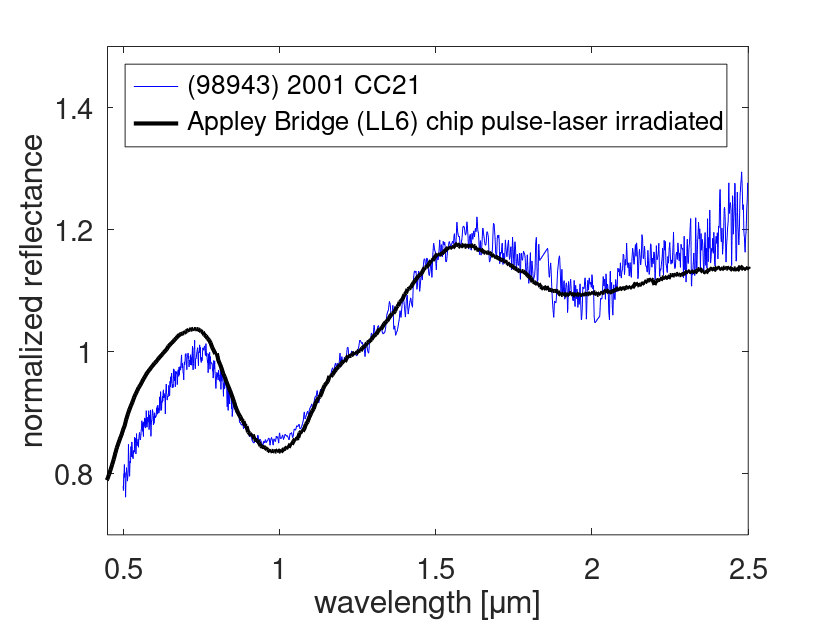}
    \caption{ The comparison between the spectrum of Torifune and the spectrum of  a sample from the meteorites Chateau Renard (L6) chip pulse-laser irradiated (Sample ID: OC-TXH-011-A80, file C1OC11A80) --left, and Appley Bridge (LL6) meteorite, a chip pulse-laser irradiated (Sample ID: OC-TXH-012-A40, file C1OC12A40) -- right . All spectra are normalized at 1.25 $\mu m$.)}
    \label{fig:meteorites}
\end{figure}

The spectrum of Torifune shows the two strong absorption features centered around 1 and 2 $\mu$m associated with silicates (mainly olivine and pyroxene). We compared this visible and near-infrared spectrum to more than 2,500 reflectance spectra of meteorites available in the RELAB database \citep{2004LPI....35.1720P}. We found that the best matches (Fig.~\ref{fig:meteorites}) correspond to samples (in the forms of chips and pellets) from Chateau Renard meteorite, an L6 ordinary chondrite irradiated with a pulse-laser (sample ID, OC-TXH-011-A80, OC-TXH-011-A60, OC-TXH-011-D35). The second best match is with a chip from Appley Bridge LL6 ordinary chondrite (Sample ID OC-TXH-012-A40), also irradiated with pulse laser. We note the evolved petrologic type (L6/LL6) of these samples. 

\begin{table}
\centering
\caption{Computed band parameters and derived modal mineralogy.}
\label{BandParam}
\begin{tabular}{l c}\hline
Parameter & Computed value\\
\hline
BImin  & 0.9559 $\pm$ 0.0087 $\mu$m \\
BIImin & 1.9962 $\pm$ 0.0073 $\mu$m \\
BIC    & 0.9911 $\pm$ 0.0087 $\mu$m \\
BIIC   & 1.9996 $\pm$ 0.0148 $\mu$m \\
%$\Delta$BIIC   & 0.013 $\mu$m \\
BAR   & 0.5196 $\pm$ 0.1629 \\
%$\Delta$BAR   & 0.053 \\
Ol/Ol+Px & 0.60 $\pm$ 0.033\\
Fa & 28.5 $\pm$ 5.1 mol \% \\
Fs & 23.4 $\pm$ 3.6 mol \% \\
\hline
\end{tabular}  
\end{table}

\begin{figure}[ht!]
\includegraphics[width=9cm]{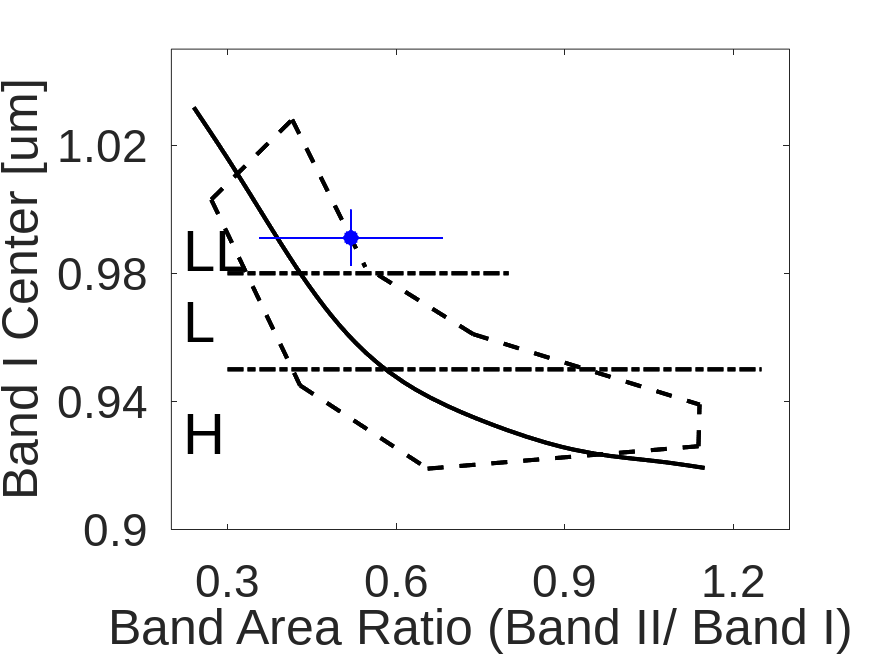}
\caption{Band Area Ratio (BAR) vs. Band I center for a sample of 48 ordinary chondrites computed by \citet{2010Icar..208..789D}. The dashed-line region encloses the obtained values. Horizontal lines separate the regions mostly occupied by H, L, and LL ordinary chordates, as labeled in the plot. The red circle corresponds to the computed BAR and Band I center values of Torifune.}
\label{BAR}
\end{figure}

The mafic mineral abundances can be derived using the two absorption features of olivine and pyroxene. We applied the models summarized by \citet{2015aste.book...43R} and reference therein \citep{1986JGR....9111641C, 1993Icar..106..573G, 2010Icar..208..789D} to compute the two band minima (BImin and BIImin), the band centers (BIC and BIIC, once the continuum has been removed) and the band area ratio (BAR) and band depths. This method was mainly developed to study main belt asteroids, and so we have to apply a temperature correction to account for the higher surface temperature of a near-Earth asteroid. We computed the temperature corrections for the Band II center ($\Delta$BIIC) and for the BAR ($\Delta$ BAR) parameter), following \cite{2012Icar..220...36S}. 

\begin{equation}
    \frac{\rm ol}{\rm ol+px}=-0.242\times BAR + 0.728~~~~~\pm0.03
    \label{olopx}
\end{equation}

\begin{equation}
    Fa=-1284.9 \times {\rm BIC}^2 + 2656.5 \times {\rm BIC} - 1342.2~~~~~\pm1.3{\rm mol}\%
    \label{Fa}
\end{equation}

\begin{equation}
    Fs=-879.1 \times {\rm BIC}^2 + 1824.9 \times {\rm BIC} -921.7~~~~~\pm1.4{\rm mol}\%
    \label{Fs}
\end{equation}

We derived the olivine/pyroxene ratio (ol/ol+px) -- Eq.~\ref{olopx}, the Fayalite (Fe$_2$SiO$_4$, Fa)  -- Eq.~\ref{Fa}  and Ferrosilite (FeSiO$_3$, Fs)  -- Eq.~\ref{Fs},  molar percentage using the relations described in \cite{2010Icar..208..789D}. The results are summarized in Table~\ref{BandParam}.

\subsection{ Estimated size}
The relation between the geometric albedo($p_V$), the effective diameter($D$) and the absolute magnitude ($H$) is given by Eq.\ref{Diameter} \citep{1997Icar..126..450H,2010Icar..209..542M}. Thus, we can use the values determined in this paper to provide an updated estimate for the effective diameter.
\begin{equation}
    D[km]=\frac{1329}{\sqrt{p_V}}\times10^{-\frac{H}{5}}
    \label{Diameter}
\end{equation}

The average albedo of Sq-type is $p_V$ = 0.243$\pm$0.039 \citep{2011ApJ...741...90M}. This value is comparable with the one reported by \cite{2011AJ....142...85T}, $p_V = 0.26^{+0.04}_{-0.03}$ for a sample of S-complex NEAs. \citet{2024A&A...688L...7F} reported  that Torifune's albedo is $p_V$ = 0.216$\pm$0.016 using the thermal data from the Spitzer Space telescope.

To propagate the associated uncertainties through Eq.~\ref{Diameter}, we generated $10^7$ clones for $H$, normally distributed around the mean value of $H = 18.78 \pm 0.14$ mag (as reported in Section 3.1), and $10^7$ clones for $p_V$, centered around its mean value of $p_V = 0.216 \pm 0.016$ (as reported by \citet{2024A&A...688L...7F}). Consequently, we found $D = 0.505 \pm 0.037$ km which is our best estimate for Torifune's effective diameter. Because \citet{2024A&A...688L...7F} reported a fainter value of $H = 18.94 \pm 0.05$, they computed a diameter of  of $D = 0.465 \pm 0.015$ km. 

\begin{figure}[ht!]
	\includegraphics[width=9cm,keepaspectratio]{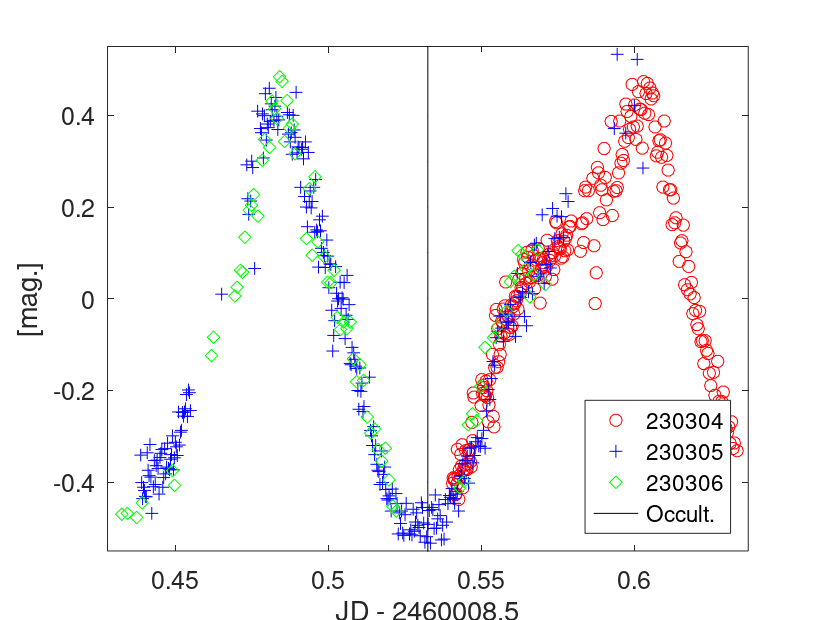}
    \caption{The extrapolated lightcurves observed around the stellar occultation event (marked by the black line) observed by citet{2024PASJ..tmp...71A}. The red circles corresponds to the observations obtained  on March 04, the blue crosses are data obtained on March 05, and the green diamonds are the photometric values obtained on March 06 2023.}
    \label{fig:arimatsu}
\end{figure}

\citet{2024PASJ..tmp...71A}  provide an estimate of the asteroid shape and size using the occultation that took place on 2023 March 05 $12^h46^m 47^s.3$ UT). Based on  Markov chain Monte Carlo (MCMC) fit they constrain the major radius $a$ = 0.42$^{+0.08}_{0.06}$ km and the ratio between the two radii  $c/a$ = $0.37 \pm 0.09$. We extrapolated the lightcurves we obtained around this date (on the night of March 04, 05, and 06, 2023) using the light-curve periodicity $P= 5.021516$ determined in previous section.  The occultation event occurred at the brightest point of the lightcurve which indicates that the major radius $a$ reported by \citet{2024PASJ..tmp...71A} is in fact the semi-major axis of Torifune.  Thus, the semi-minor axis $c$ = 0.16$^{+0.05}_{0.04}$ km. The third semi-axis can be estimated from the light-curve amplitude which is 0.96 mag. resulting in $b/a$ = 0.41, consequently $b$ = $0.17\pm0.03$ km. Based on these values we can compute the volume equivalent diameter $D_{eq}$ = $0.44 \pm 0.06$ km. The uncertainties are obtained by propagating the corresponding errors.

To explain the discrepancy between the ${H,G}$ model and the observation obtained at aspect angles of around 50$^\circ$, we compute the polar-to-equatorial oblateness as $R = \frac{c(a + b)}{2ab} = 0.68 \pm 0.24$. Using Eq.\ref{deltamag}, we obtain a magnitude difference of $\Delta_{mag}^\Theta = -0.30 \pm 0.25$ mag for $\Theta = 50 \pm 20^\circ$. This result is consistent with the observations shown in Fig.~\ref{fig:HG} and may help explain the difference compared to the $H$ value reported by \citet{2024A&A...688L...7F}.

We note that the volume equivalent diameter $D_{eq}$ = $0.44 \pm 0.06$ km is  smaller compared to the diameter computed from the absolute magnitude $D = 0.505 \pm 0.037$. This is because the absolute magnitude determination depends on the surface seen by the observer, and not by the volume. 

\section{Discussion} \label{sec:discuss}

We conducted a series of photometric and spectroscopic observations of Torifune, the target for the Hayabusa2$\#$ flyby. Our findings indicate that Torifune is an oblate object with no detectable surface heterogeneity. Our spectroscopic data from the GTC and IRTF clearly classify this object as Sq type. The spectral curve is in agreement with those reported by \citet{2023MNRAS.525L..17G} and \citet{2005MNRAS.359.1575L}, and discard a possible L-type classification.

%The Sq-type spectral curve shows 1 and 2 micron absorption bands characteristic to olivine and pyroxene compositions. This is an intermediate type between the Q and S types, which can be the space weathering track . Alternatively, this could be also interpreted as the correlation with LL to H ordinary chondrites \citep{DeMeo2022}. 

Our mineralogical analysis indicates that Torifune exhibits spectral characteristics consistent with LL/L ordinary chondrites (Fig.~\ref{BAR}). This result is further supported by the direct comparison with the laboratory spectra of meteorites. The best spectral matches are found with  with the samples of ordinary chondirtes, specifically the L6  Ch{\^a}teau Renard and the LL6 Appley Bridge.  The Fig.~\ref{fig:meteorites}  illustrates the curve fit of band shapes and spectral features  (e.g. the shoulder at 1.3~$\mu m$) between the meteorites and Torifune spectra.  The discrepancies are mostly in the visible region (when the two spectra are normalized at 1.25~$\mu$m), and can be attributed to differences in spectral slope.

The spectral matching with meteorites (with the top four best matches corresponding to pulse laser-irradiated samples) suggests that Torifune has a space-weathered surface, as both laboratory samples were irradiated with a pulse laser \citep{2006Icar..184..327B}. Space weathering is the process that alters the optical properties of the surfaces of airless rocky bodies due to exposure to the space environment \citep{2012MNRAS.421....2M}. This finding is further supported by the classification of Torifune as an Sq-type asteroid. Traditionally, Q-type asteroids are interpreted as having fresh ordinary-chondrite-like compositions, S-types are thought to have space-weathered surfaces \citep[e.g.,][]{2002aste.book..585C,2010Natur.463..331B,2024M&PS...59.1329M}, and Sq-types are considered to have moderately space-weathered surfaces \citep[e.g.,][]{Sasaki2001,2004M&PS...39..351B,Koga2018}.

The Relab database contains spectra of multiple samples from  Ch{\^a}teau Renard (8 samples) and Appley Bridge (3 samples). These samples, both fresh and irradiated with various laser energies, are detailed in \citet{2024M&PS...59.1329M}. Notably, the best spectral matches between the Torifune spectrum and meteorites spectral curves are observed with those samples irradiated with the highest laser energy. Spectral reddening and darkening are changes observed in all experiments simulating space-weathering effects on silicates and meteorites \citep{2015aste.book..597B}. These experiments also demonstrate that space weathering does not alter the band centers or the band area ratio. However, \citet{2012MNRAS.421....2M} observed that the spectral reddening of asteroids results from a combination of three factors: exposure to space weathering over the object's lifetime, its original composition, and the structure and texture of its surface.

\begin{figure}[ht!]
	\includegraphics[width=9cm,keepaspectratio]{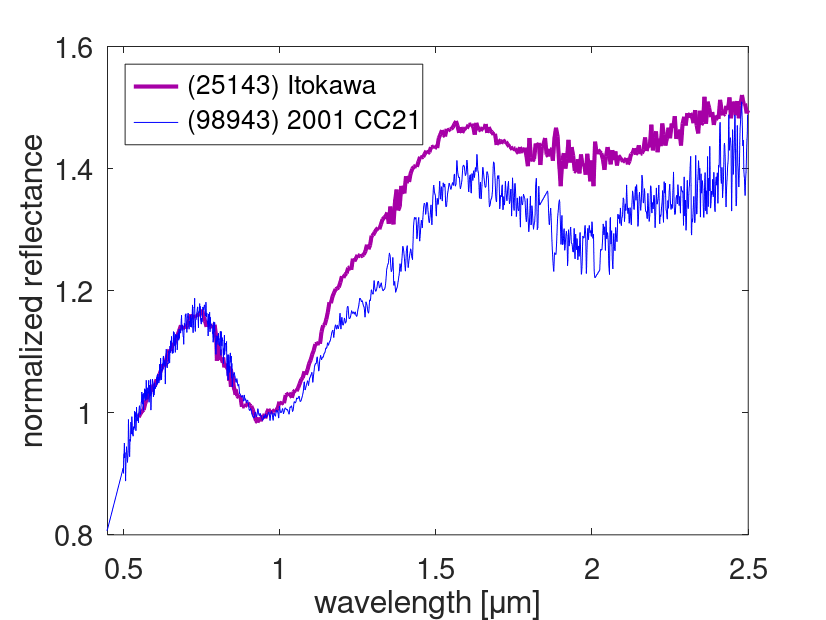}
    \caption{ The comparison between the spectrum of (99843) Torifune and (25143) Itokawa \citep{2001M&PS...36.1167B}. All spectra are normalized at 0.55 $\mu$.)}
    \label{fig:Itokawa}
\end{figure}

Interestingly, Torifune has several common points, such as albedo, reflectance spectrum (Fig.~\ref{fig:Itokawa}), size, and shape, with (25143) Itokawa, visited by the Hayabusa spacecraft in 2006 \citep{Fujiwara2006}. Itokawa is a rubble-pile elongated asteroid, which is classified as an Sq-type asteroid, with the  most likely composition similar with LL chondrites \citep{Binzel2001}. If (98943) Torifune is similar to Itokawa, we expect to see quite large spectral variation on its surface ranging from Q to S type \citep{Koga2018}. The ONC-T has narrow-band filters ranging from 0.4 to 0.95 $\mu$m \citep{Kameda2017, Tatsumi2019}, which can detect the difference in the 1-$\mu$m absorption. In fact, the ONC-T has detected the mafic silicates on the surface of the carbonaceous asteroid (162173) Ryugu \citep{Tatsumi2021}.

The differences between  the spectrum of (98943) Torifune and  of (25143) Itokawa \citep{2001M&PS...36.1167B}, observed with ground-based telescopes are mostly due to a different spectral slope. This suggests a different space-weathering degree. Both of these curves represent the disk-integrated spectra. The in-situ observations made by Haybusa  to Itokawa shown spectral differences across the object which are consistent with the space-weathering trend \citep{2006Natur.443...56H}.

\section{Conclusions} \label{sec:discuss}
The near-Earth asteroid (98943) Torifune, formally called 2001 CC$_{21}$, is the flyby target of Hayabusa2 extended mission, nicknamed Hayabusa2 $\#$. We obtained a detailed characterization of it during 2022 and 2023 when this object had an apparent magnitude as bright as 16.5. 
\begin{itemize}
    \item We obtained its absolute magnitude $H~=~18.78~\pm~0.14$ mag and its rotation period $P~=~5.021516\pm0.000106$ h.
    \item We determined its convex shape and its pole solution $\lambda = 301^{\circ} \pm 35^{\circ}$, $\beta = 89^{\circ} \pm 6^{\circ}$ and $\epsilon = 5^{\circ} \pm 3^{\circ}$. These imply a prograde rotator.
    \item We also estimated the semi-axis of the equivalent ellipsoid, $a$ = 0.42$^{+0.08}_{0.06}$ km , $b$ = 0.16$^{+0.05}_{0.04}$ km,  and $c$ = $0.17\pm0.03$ km by comparing our lightcurve results with the occultation data reported by \citet{2024PASJ..tmp...71A}.  Consequently, the volume equivalent diameter is $D_{eq}$ = $0.44 \pm 0.06$ km .     
    \item We found that (98943) Torifune exhibits no large-scale heterogeneity, based on simultaneous observations performed with $g$, $r$, $i$, and $z_s$ filters.
    \item We report the average colors $(g-r)~=~0.663\pm 0.022$ mag, $(r-i)~=~0.177\pm 0.012$ mag, $(i-z_s)~=~-0.061\pm0.032$ mag,	and $(r-z_s)~=~0.116\pm0.025$ mag which indicate an S-complex classification.  Using the INT telescope we also found the $(B-V)$ = 1.142$\pm$0.09 mag, and $(V-R)$ = 0.461$\pm$0.05 mag colors.
    \item We spectrally classified it as an Sq-type in Bus-DeMeo taxonomy, and we infer a mineralogy similar to L6/LL6 ordinary chondrites with  a moderate space-weathered surface.
    \item We determined the major compositional constituents: ol/(ol+px) = 0.60, a Fa content of 28.5 mol$\%$, and an Fs content of 23.4 mol$\%$.
    \item  The differences between the disk-integrated spectra of (98943) Torifune, presented in this study, and (25143) Itokawa \citep{2001M&PS...36.1167B} are primarily attributed to variations in the spectral slope, indicating differing degrees of space weathering. This suggests a similar average composition for both objects, with Torifune exhibiting more pronounced space weathering effects compared to those observed by Hayabusa on Itokawa \citep{2006Natur.443...56H}.    
\end{itemize}

\begin{acknowledgments}
The work of MP and DB was supported by a grant from the Romanian National Authority for  Scientific  Research -- UEFISCDI, project number PN-III-P2-2.1-PED-2021-3625. This work is supported by JAXA Hayabusa2 extended project. MSR used flash storage and GPU computing resources as Indefeasible Computer Rights (ICRs) being commissioned at the ASTRO POC project that Light Bridges will operate in the Island of Tenerife, Canary Islands (Spain). The work of JFM was supported by the Gobierno de Espa\~{n}a, Ministerio de Econom\'{i}a, Comercio y Empresa, ICEX Vives program, grant number P1559. The ICRs used for this research were provided by Light Bridges in cooperation with Hewlett Packard Enterprise (HPE) and VAST DATA. This article includes observations made with the Two-meter Twin Telescope (TTT, TTT1-Y65 and TTT2-Y66) sited at the Teide Observatory of the IAC that Light Bridges operates on the Island of Tenerife, Canary Islands (Spain). The Observing Time Rights (DTO) used for this research were provided by IAC. Based on observations made with the Gran Telescopio Canarias (GTC), programme number GTC32-22B, installed at the Spanish Observatorio del Roque de los Muchachos of the Instituto de Astrof\'{\i}sica de Canarias, on the island of La Palma. Based on observations made with  Carlos S{\'a}nchez Telescope at the IAC’s Teide Observatory (program ID OOT2022-370). Based on observations made with the Isaac Newton Telescope (INT), in the Spanish Observatorio del Roque de los Muchachos of the Instituto de Astrof\'{\i}sica de Canarias (program ID SI2022a03), observations made by dr. Ovidiu Vaduvescu and Rosa Hoogenboom. Based on observations made with  NASA Infrared Telescope Facility SpeX --  program ID  2023A066,  which is operated by the University of Hawaii under contract 80HQTR24DA010 with the National Aeronautics and Space Administration.This research utilizes spectra acquired by Takahiro Hiroi with the NASA RELAB facility at Brown University.
\end{acknowledgments}

%% To help institutions obtain information on the effectiveness of their 
%% telescopes the AAS Journals has created a group of keywords for telescope 
%% facilities.
%
%% Following the acknowledgments section, use the following syntax and the
%% \facility{} or \facilities{} macros to list the keywords of facilities used 
%% in the research for the paper.  Each keyword is check against the master 
%% list during copy editing.  Individual instruments can be provided in 
%% parentheses, after the keyword, but they are not verified.

\vspace{5mm}
\facilities{
Two-meter Twin Telescope facility (TTT),
Carlos S{\'a}nchez Telescope (TCS),
Gran Telescopio Canarias (GTC),
Infrared Telescope Facility (NASA/IRTF),
Instituto de Astrofisica de Canarias(IAC80),
Isaac Newton Telescope (INT),
T025 - BD4SB telescope}

%% Similar to \facility{}, there is the optional \software command to allow 
%% authors a place to specify which programs were used during the creation of 
%% the manuscript. Authors should list each code and include either a
%% citation or url to the code inside ()s when available.

\software{astropy \citep{2013A&A...558A..33A,2018AJ....156..123A},  
          Source Extractor \citep{1996A&AS..117..393B},
          IRAF \citep{1986SPIE..627..733T},
          GNU Astronomy Utilities \citep{gnuastro},
          GNU Octave \citep{octave}
          Cloudy \citep{2013RMxAA..49..137F}, 
          }

%% Appendix material should be preceded with a single \appendix command.
%% There should be a \section command for each appendix. Mark appendix
%% subsections with the same markup you use in the main body of the paper.

%% Each Appendix (indicated with \section) will be lettered A, B, C, etc.
%% The equation counter will reset when it encounters the \appendix
%% command and will number appendix equations (A1), (A2), etc. The
%% Figure and Table counter will not reset.

\appendix

\section{Appendix information}

%Appendices can be broken into separate sections just like in the main text.
%% For this sample we use BibTeX plus aasjournals.bst to generate the
%% the bibliography. The sample631.bib file was populated from ADS. To
%% get the citations to show in the compiled file do the following:
%%
%% pdflatex sample631.tex
%% bibtext sample631
%% pdflatex sample631.tex
%% pdflatex sample631.tex

%\begin{longrotatetable}
%\begin{longtable}
\startlongtable
\begin{deluxetable*}{lcccccccll}
\label{photomlogtable}
\tablecaption{Observing Log. The  number of acquired images (N.imags), the time for the start ($UT_{start}$) and end ($UT_{end}$) of the observations, the heliocentric distance ($r$) expressed in au, the geocentric distance ($\Delta$) in au, the phase angle ($\alpha$) in $^\circ$, the predicted apparent V magnitude ($Vmag$), the filter, and the instrument (camera and telescope) used are provided. The ephmerides were obtained using the Horizons System - JPL Solar System Dynamics, accessed on December 1, 2024.}
\tablewidth{700pt}
%\tabletypesize{\scriptsize}
\tablehead{
\colhead{Date obs.} & \colhead{N.imags} & 
\colhead{$UT_{start}$} & \colhead{$UT_{end}$} & 
\colhead{$r$} & \colhead{$\Delta$} & 
\colhead{$\alpha$} & \colhead{$Vmag$} & \colhead{filter} & \colhead{Tel}
} 
\startdata
2023-02-08 & 361  & 22:32 & 03:20 & 1.09538 & 0.14250 & 37.7 & 16.2 & Lum & QHY411-TTT2\\
2023-02-12 & 496  & 21:14 & 03:32 & 1.08183 & 0.13846 & 44.2 & 16.3 & Lum & QHY411-TTT2\\
2023-02-13 & 176  & 22:03 & 04:16 & 1.07825 & 0.13762 & 45.9 & 16.3 & Lum & QHY411-TTT2\\
2023-02-14 & 204  & 02:03 & 03:28 & 1.07453 & 0.13685 & 47.7 & 16.4 & Lum & QHY411-TTT2\\
2023-02-18 & 347  & 00:00 & 04:01 & 1.06046 & 0.13461 & 54.5 & 16.5 & Lum & QHY411-TTT2\\
2023-02-27 & 489  & 21:49 & 01:16 & 1.02789 & 0.13211 & 70.0 & 16.8 & Lum & iKon936-TTT1\\
2023-02-28 & 879  & 20:28 & 01:59 & 1.02420 & 0.13195 & 71.7 & 16.9 & Lum & iKon936-TTT1\\
2023-03-01 & 569  & 21:36 & 01:34 & 1.02040 & 0.13180 & 73.4 & 16.9 & Lum & iKon936-TTT1\\
2023-03-02 & 20   & 20:34 & 00:05 & 1.01685 & 0.13167 & 75.1 & 17.0 & Lum & QHY411-TTT2\\
2023-03-02 & 272  & 20:33 & 23:31 & 1.01689 & 0.13167 & 75.0 & 17.0 & Lum & iKon936-TTT1\\
2023-03-03 & 450  & 20:42 & 01:36 & 1.01295 & 0.13153 & 76.9 & 17.0 & Lum & iKon936-TTT1\\
2023-03-04 & 317  & 21:55 & 01:06 & 1.00912 & 0.13140 & 78.6 & 17.1 & Lum & iKon936-TTT1\\
2023-03-05 & 296  & 20:35 & 00:31 & 1.00549 & 0.13127 & 80.3 & 17.1 & Lum & iKon936-TTT1\\
2023-03-06 & 111  & 20:46 & 00:52 & 1.00166 & 0.13115 & 82.1 & 17.2 & Lum & iKon936-TTT1\\
2023-11-22 & 253  & 01:48 & 04:58 & 1.25816 & 0.35513 & 34.8 & 18.4 & g & QHY411-TTT2\\
2023-11-23 & 131  & 02:58 & 05:48 & 1.25833 & 0.35029 & 34.0 & 18.4 & g & QHY411-TTT2\\
2023-11-27 & 375  & 00:33 & 06:03 & 1.25859 & 0.33262 & 30.6 & 18.2 & g & QHY411-TTT2\\
2023-11-28 & 933  & 00:02 & 05:51 & 1.25855 & 0.32841 & 29.6 & 18.1 & g & QHY411-TTT2\\
2023-12-01 & 573  & 23:16 & 06:00 & 1.25820 & 0.31618 & 26.7 & 17.9 & g & QHY411-TTT2\\
2023-12-02 & 896  & 23:45 & 05:58 & 1.25800 & 0.31224 & 25.7 & 17.9 & g & QHY411-TTT2\\
2023-12-03 & 1355 & 22:42 & 02:16 & 1.25778 & 0.30882 & 24.7 & 17.8 & g & QHY411-TTT2\\
2023-12-07 & 311  & 03:27 & 05:10 & 1.25636 & 0.29424 & 20.1 & 17.6 & g & QHY411-TTT2\\
2023-12-08 & 904  & 00:03 & 05:43 & 1.25594 & 0.29124 & 19.0 & 17.5 & g & QHY411-TTT2\\
2023-12-09 & 938  & 22:52 & 06:20 & 1.25546 & 0.28823 & 17.8 & 17.5 & g & QHY411-TTT2\\
2023-12-11 & 384  & 03:08 & 05:28 & 1.25432 & 0.28237 & 15.3 & 17.3 & g & QHY411-TTT2\\
2023-12-12 & 264  & 04:35 & 06:40 & 1.25368 & 0.27964 & 13.9 & 17.2 & g & QHY411-TTT1\\
2023-12-13 & 248  & 04:02 & 06:36 & 1.25303 & 0.27723 & 12.7 & 17.2 & g & QHY411-TTT2\\
2023-12-14 & 1128 & 21:53 & 05:48 & 1.25244 & 0.27526 & 11.6 & 17.1 & g & QHY411-TTT2\\
2023-12-14 & 787  & 02:52 & 06:38 & 1.25235 & 0.27500 & 11.5 & 17.1 & g & QHY411-TTT1\\
2023-12-15 & 112  & 22:15 & 06:23 & 1.25170 & 0.27308 & 10.4 & 17.1 & g,r,i,$z_s$ & iKon936-TTT1\\
2023-12-16 & 151  & 03:07 & 05:55 & 1.25085 & 0.27095 & 9.0  & 17.0 & g,r,i & QHY411-TTT2\\
2023-12-17 & 1109 & 22:20 & 05:58 & 1.25012 & 0.26932 & 8.0  & 16.9 & g & iKon936-TTT1\\
2023-12-18 & 459  & 02:49 & 06:50 & 1.24917 & 0.26753 & 6.7  & 16.9 & g & QHY411-TTT2\\
2023-12-19 & 932  & 20:05 & 06:24 & 1.24841 & 0.26628 & 5.9  & 16.8 & g & QHY411-TTT2\\
2023-12-20 & 435  & 00:00 & 05:44 & 1.24742 & 0.26491 & 5.1  & 16.8 & g & QHY411-TTT2\\
2023-12-25 & 1446 & 19:41 & 06:01 & 1.24225 & 0.26113 & 6.9  & 16.8 & g,r,i,$z_s$& QHY411-TTT2\\
2023-12-25 & 30   & 22:23 & 23:57 & 1.24233 & 0.26116 & 6.9  & 16.8 & g & iKon936-TTT1\\
2023-12-26 & 1385 & 19:42 & 04:05 & 1.24112 & 0.26088 & 8.0  & 16.8 & g & QHY411-TTT2\\
2023-12-27 & 1258 & 19:30 & 05:11 & 1.23989 & 0.26079 & 9.3  & 16.9 & g & QHY411-TTT2\\
2023-12-28 & 1832 & 19:43 & 05:35 & 1.23862 & 0.26087 & 10.6 & 16.9 & g & QHY411-TTT2\\
2023-12-29 & 2045 & 19:43 & 05:29 & 1.23734 & 0.26111 & 12.0 & 17.0 & g & QHY411-TTT2\\
2023-12-30 & 1390 & 20:38 & 04:21 & 1.23601 & 0.26152 & 13.3 & 17.1 & g & QHY411-TTT2\\
2024-01-02 & 484  & 00:13 & 03:22 & 1.23170 & 0.26375 & 17.5 & 17.2 & g & QHY411-TTT2\\
2024-01-03 & 225  & 00:12 & 02:02 & 1.23026 & 0.26475 & 18.8 & 17.3 & g & QHY411-TTT2\\
2024-01-04 & 429  & 23:52 & 02:37 & 1.22872 & 0.26593 & 20.2 & 17.3 & g & QHY411-TTT2\\
2024-01-08 & 137  & 23:09 & 01:22 & 1.22228 & 0.27190 & 25.3 & 17.5 & g & iKon936-TTT1\\
2024-01-09 & 127  & 23:18 & 00:13 & 1.22059 & 0.27368 & 26.6 & 17.5 & g & iKon936-TTT1\\
2024-01-19 & 359  & 22:18 & 02:46 & 1.20120 & 0.29710 & 37.9 & 18.0 & g & QHY411-TTT2\\
2024-01-20 & 312  & 22:48 & 03:11 & 1.19902 & 0.29987 & 38.9 & 18.1 & g & QHY411-TTT2\\
2023-02-08 & 148  & 21:34 & 02:08 & 1.09554 & 0.14256 & 37.6 & 16.2 & Clear & T025-BD4SB\\
2023-02-09 & 260  & 18:18 & 00:41 & 1.09250 & 0.14152 & 39.1 & 16.2 & Clear & T025-BD4SB\\
2023-02-14 & 591  & 18:33 & 01:52 & 1.07519 & 0.13697 & 47.4 & 16.4 & Clear & T025-BD4SB\\
2023-02-15 & 524  & 18:16 & 02:09 & 1.07168 & 0.13630 & 49.1 & 16.4 & Clear & T025-BD4SB\\
2023-02-16 & 229  & 18:53 & 00:43 & 1.06821 & 0.13571 & 50.8 & 16.4 & Clear & T025-BD4SB\\
2022-11-25 & 85   & 02:16 & 06:28 & 1.25283 & 0.49385 & 47.3 & 19.5 & $r'$ & IAC80\\
2022-12-20 & 230  & 00:28 & 07:00 & 1.22350 & 0.33435 & 38.5 & 18.3 & $r'$ & IAC80\\
2023-02-27 & 283  & 20:17 & 23:59 & 1.02811 & 0.13212 & 69.8 & 16.8 & $r'$ & IAC80\\
2023-03-01 & 163  & 20:50 & 23:03 & 1.02066 & 0.13181 & 73.3 & 16.9 & $r'$ & IAC80\\
\enddata
\end{deluxetable*}
%\end{longrotatetable}
%\end{longtable}

\startlongtable
\begin{deluxetable*}{lccccc}
\label{photomlogtable2}
\tablecaption{Observing geometry. The average UT time of the observations, the phase angle ($\alpha$ [$^\circ$]), the average observed apparent magnitude $V$, the average reduced V magnitude ($V_{red}$), the aspect angle $\Theta[^\circ]$ and its uncertainity $\sigma_{\Theta}[^\circ]$.}
\tablewidth{700pt}
%\tabletypesize{\scriptsize}
\tablehead{
\colhead{$UT_{avg}$} & \colhead{$\alpha$}  & 
\colhead{$V$} & \colhead{$V_{red}$} & 
\colhead{$\Theta$} & \colhead{$\sigma_{\Theta}$}
} 
\startdata
2023-02-13T00:23 & 44.2 & 16.30 & 20.42 & 55 & 18 \\
2023-02-14T01:09 & 45.9 & 16.38 & 20.53 & 54 & 18 \\
2023-02-19T02:00 & 54.5 & 16.50 & 20.72 & 53 & 19 \\
2023-02-27T23:32 & 70.0 & 17.10 & 21.44 & 51 & 21 \\
2023-02-28T23:13 & 71.7 & 17.08 & 21.42 & 51 & 21 \\
2023-03-01T23:35 & 73.4 & 17.18 & 21.54 & 51 & 20 \\
2023-03-02T22:19 & 75.1 & 17.20 & 21.57 & 51 & 20 \\
2023-03-02T22:02 & 75.0 & 17.32 & 21.69 & 51 & 20 \\
2023-03-03T23:09 & 76.9 & 17.25 & 21.62 & 51 & 19 \\
2023-03-04T23:30 & 78.6 & 17.33 & 21.72 & 51 & 20 \\
2023-03-05T22:33 & 80.3 & 17.37 & 21.77 & 51 & 20 \\
2023-03-06T22:49 & 82.1 & 17.58 & 21.99 & 51 & 19 \\
2023-11-23T03:23 & 34.8 & 18.27 & 20.02 & 92 & 20 \\
2023-11-24T04:23 & 34.0 & 18.45 & 20.23 & 92 & 20 \\
2023-11-28T03:18 & 30.6 & 17.91 & 19.80 & 91 & 20 \\
2023-11-29T02:56 & 29.6 & 17.86 & 19.77 & 91 & 20 \\
2023-12-02T02:38 & 26.7 & 17.88 & 19.89 & 90 & 21 \\
2023-12-03T02:51 & 25.7 & 17.83 & 19.86 & 90 & 21 \\
2023-12-04T00:29 & 24.7 & 17.73 & 19.78 & 90 & 21 \\
2023-12-08T04:18 & 20.1 & 17.41 & 19.57 & 89 & 20 \\
2023-12-09T02:53 & 19.0 & 17.56 & 19.75 & 89 & 21 \\
2023-12-10T02:36 & 17.8 & 17.40 & 19.61 & 88 & 21 \\
2023-12-12T04:18 & 15.3 & 17.37 & 19.62 & 88 & 22 \\
2023-12-13T05:37 & 13.9 & 17.08 & 19.36 & 88 & 21 \\
2023-12-14T05:19 & 12.7 & 17.19 & 19.48 & 87 & 21 \\
2023-12-15T01:50 & 11.6 & 17.19 & 19.50 & 87 & 21 \\
2023-12-15T04:45 & 11.5 & 17.19 & 19.51 & 87 & 21 \\
2023-12-16T02:19 & 10.4 & 17.12 & 19.46 & 87 & 21 \\
2023-12-17T04:31 & 9.0 & 16.97 & 19.32 & 86 & 21  \\
2023-12-18T02:09 & 8.0 & 16.97 & 19.33 & 86 & 21  \\
2023-12-19T04:49 & 6.7 & 16.85 & 19.23 & 86 & 21  \\
2023-12-20T01:14 & 5.9 & 16.84 & 19.24 & 86 & 21  \\
2023-12-21T02:52 & 5.1 & 16.83 & 19.23 & 85 & 21  \\
2023-12-26T00:51 & 6.9 & 16.78 & 19.23 & 84 & 21  \\
2023-12-25T23:10 & 6.9 & 16.85 & 19.29 & 84 & 21  \\
2023-12-26T23:53 & 8.0 & 16.81 & 19.26 & 84 & 21  \\
2023-12-28T00:20 & 9.3 & 16.89 & 19.35 & 84 & 21  \\
2023-12-29T00:39 & 10.6 & 16.95 & 19.40 & 83 & 22 \\
2023-12-30T00:36 & 12.0 & 16.96 & 19.41 & 83 & 21 \\
2023-12-31T00:29 & 13.3 & 17.01 & 19.46 & 83 & 21 \\
2024-01-03T01:47 & 17.5 & 17.24 & 19.68 & 82 & 21 \\
2024-01-04T01:07 & 18.8 & 17.18 & 19.62 & 82 & 21 \\
2024-01-05T01:14 & 20.2 & 17.27 & 19.70 & 82 & 21 \\
2024-01-09T00:15 & 25.3 & 17.42 & 19.81 & 81 & 21 \\
2024-01-09T23:45 & 26.6 & 17.40 & 19.78 & 81 & 21 \\
2024-01-20T00:32 & 37.9 & 17.95 & 20.19 & 80 & 20 \\
2024-01-21T00:59 & 38.9 & 17.96 & 20.18 & 80 & 20 \\
2023-02-08T23:51 & 37.6 & 16.08 & 20.11 & 57 & 17 \\
2023-02-09T21:29 & 39.1 & 16.03 & 20.09 & 56 & 17 \\
2023-02-14T22:12 & 47.4 & 16.28 & 20.44 & 54 & 19 \\
2023-02-15T22:12 & 49.1 & 16.39 & 20.57 & 54 & 19 \\
2023-02-16T21:48 & 50.8 & 16.29 & 20.49 & 53 & 19 \\
2022-11-26T04:22 & 47.3 & 19.36 & 20.40 & 89 & 15 \\
2022-12-21T03:44 & 38.5 & 18.30 & 20.24 & 83 & 12 \\
2023-02-27T22:08 & 69.8 & 16.99 & 21.32 & 51 & 21 \\
2023-03-01T21:56 & 73.3 & 17.14 & 21.49 & 51 & 21 \\
\enddata
\end{deluxetable*}
%\end{longrotatetable}
%\end{longtable}

\section{Model Fit}
\begin{figure*}
\centering
        \includegraphics[width=4cm]{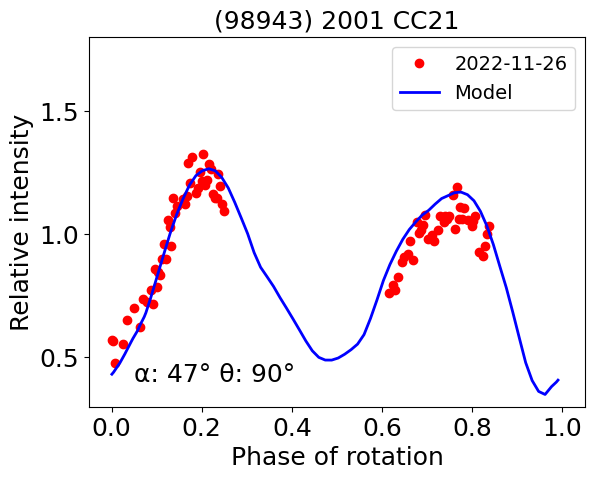}
        \includegraphics[width=4cm]{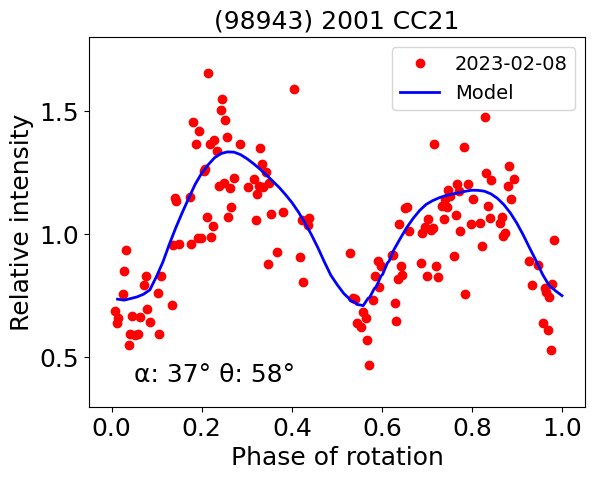}
        \includegraphics[width=4cm]{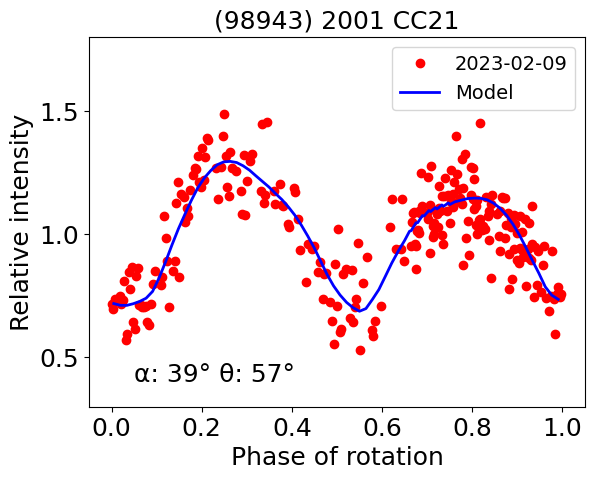}
        \includegraphics[width=4cm]{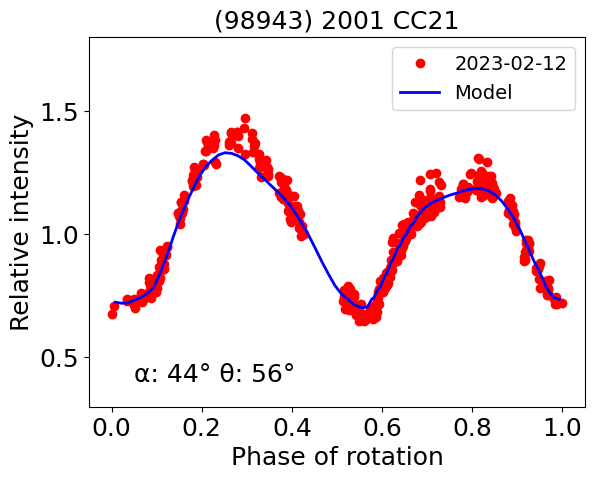}
        \includegraphics[width=4cm]{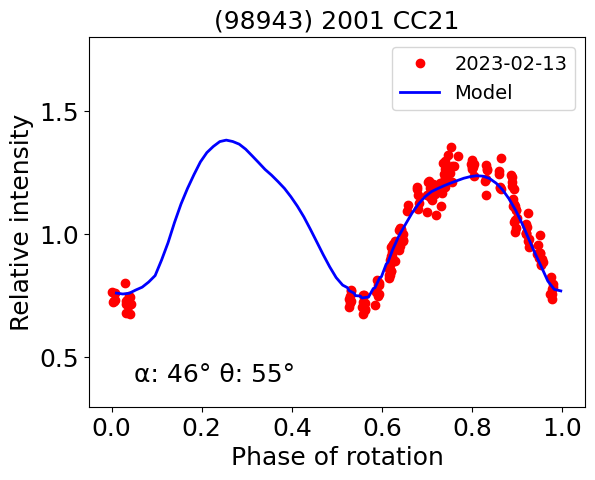}
        \includegraphics[width=4cm]{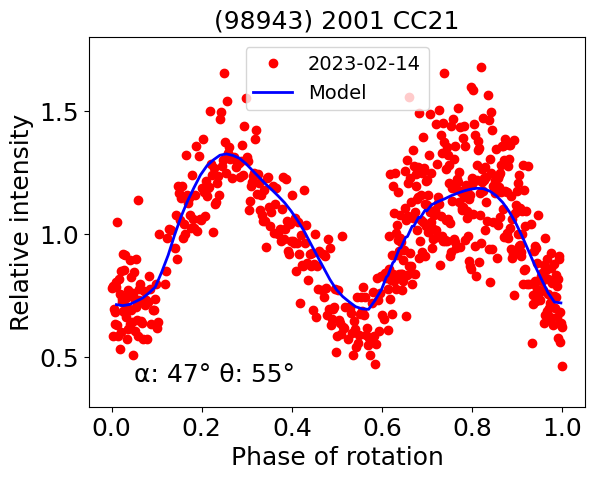}
        \includegraphics[width=4cm]{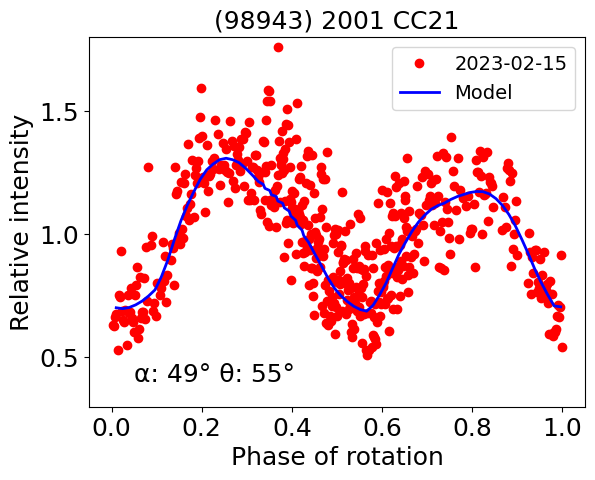}
        \includegraphics[width=4cm]{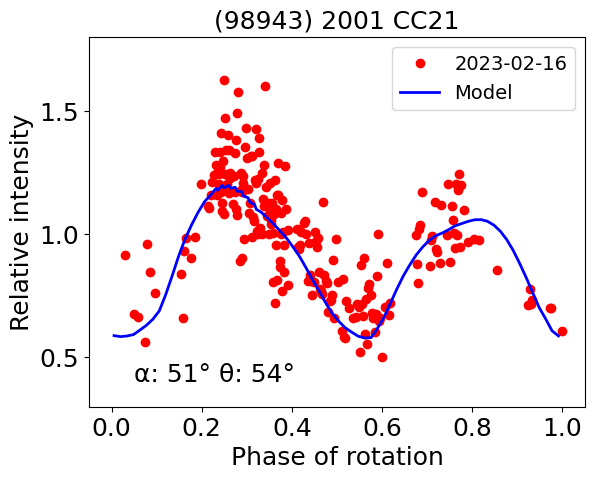}
        \includegraphics[width=4cm]{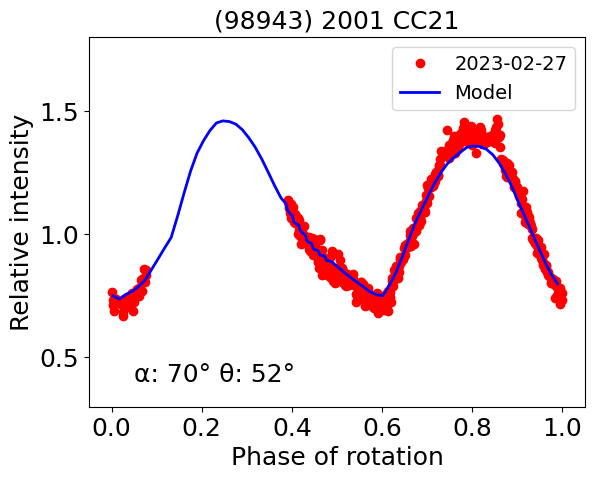}
        \includegraphics[width=4cm]{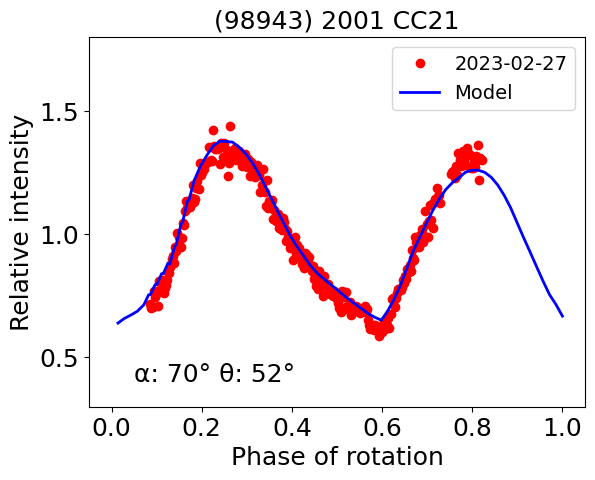}
        \includegraphics[width=4cm]{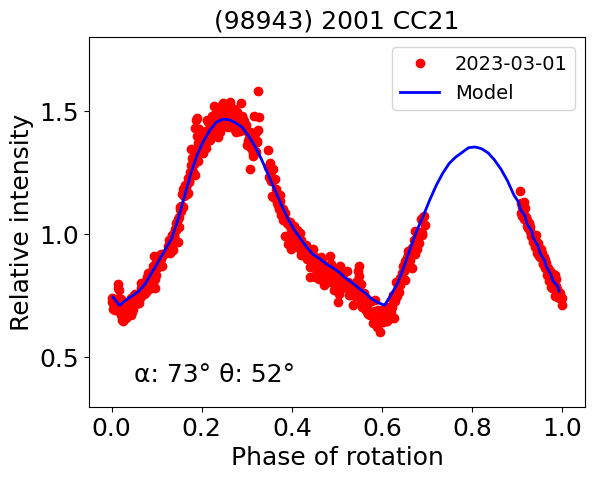}
        \includegraphics[width=4cm]{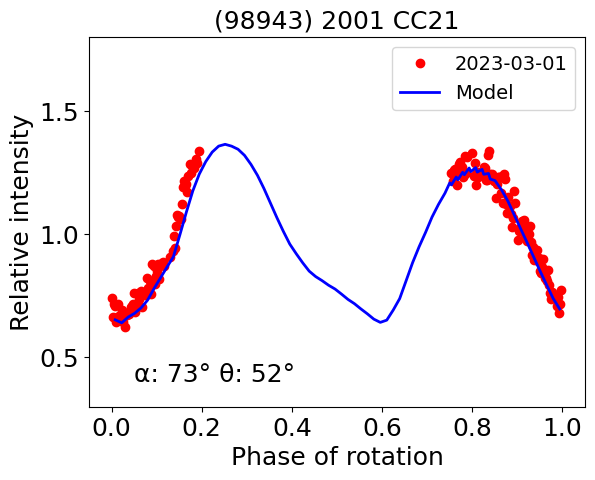}
        \includegraphics[width=4cm]{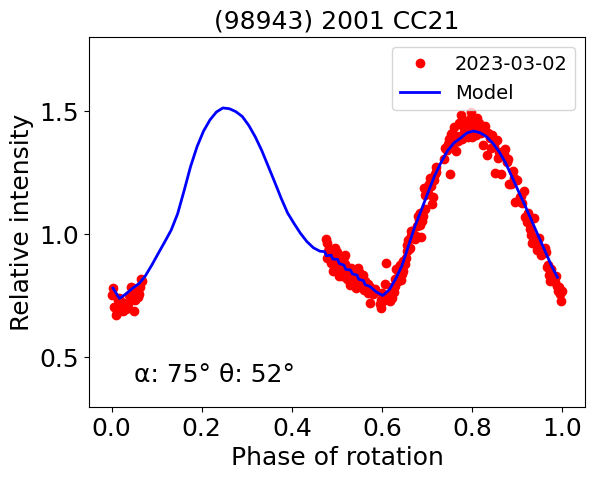}
        \includegraphics[width=4cm]{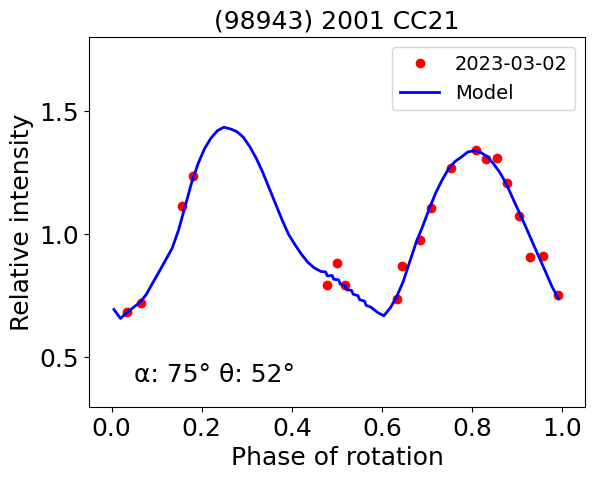}
        \includegraphics[width=4cm]{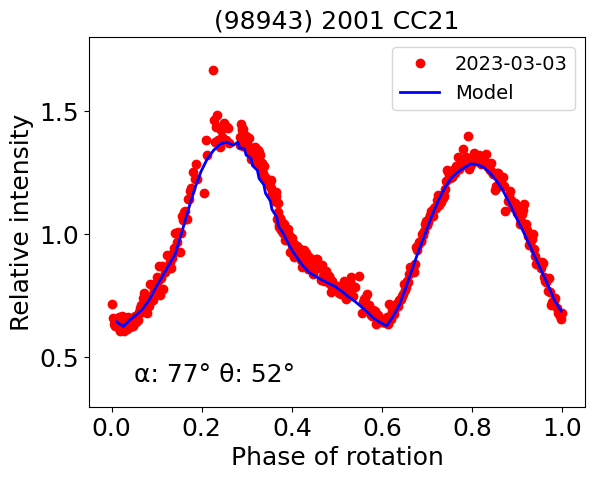}
        \includegraphics[width=4cm]{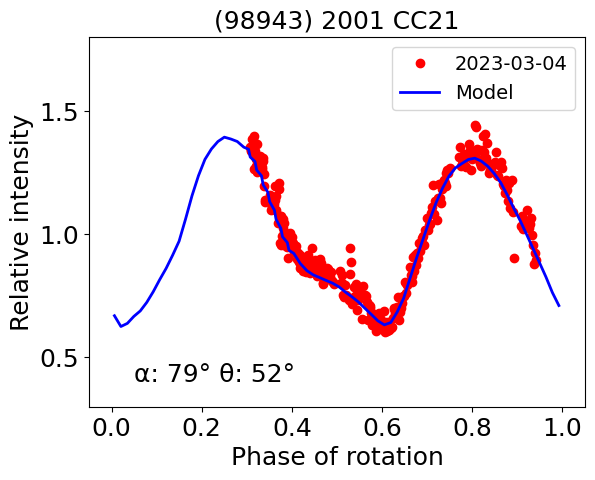}
        \includegraphics[width=4cm]{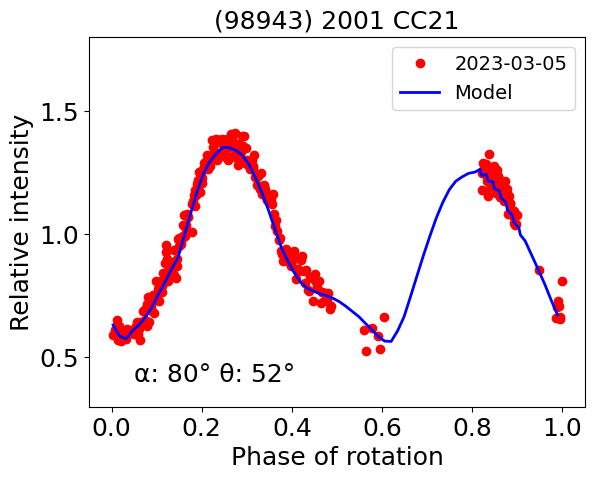}
        \includegraphics[width=4cm]{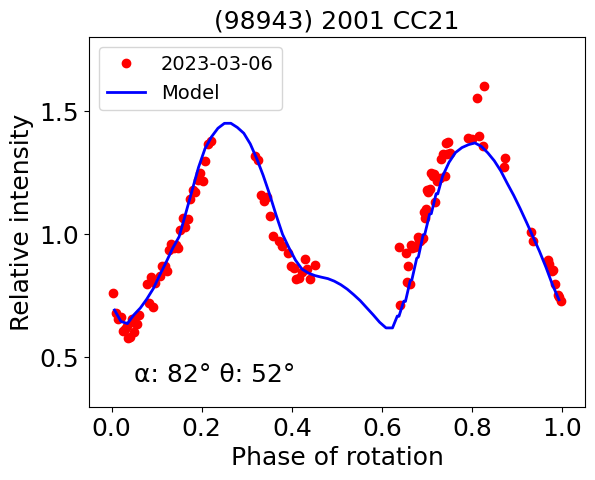}
        \includegraphics[width=4cm]{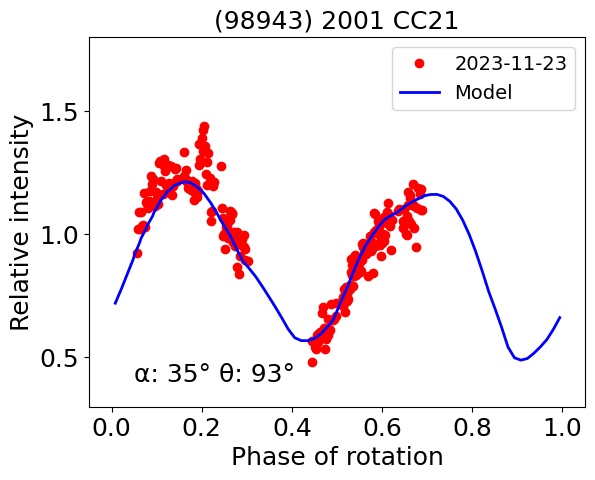}
        \includegraphics[width=4cm]{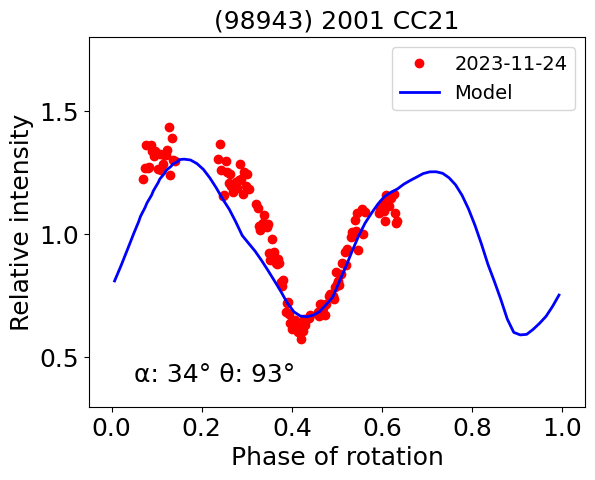}
    \caption{Graphical representation of the fit between the rest of the light curves and the best-obtained model for (98943) Torifune. The observed data is plotted as red dots, while the shape model is plotted as a solid blue line for each of the observations. The geometry is described by its solar phase angle $\alpha$ and its aspect angle $\theta$. The reference Julian Day is 2459909.000000.}
    \label{fig:lc_fit_rest}
\end{figure*}

\addtocounter{figure}{-1}
\begin{figure*}
\centering
    
        \includegraphics[width=4cm]{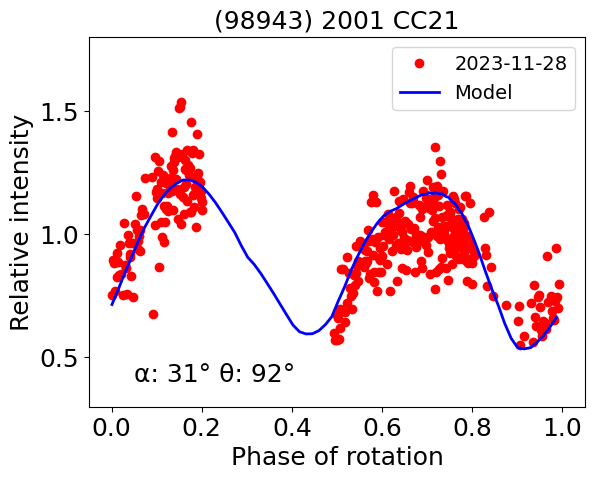}
        \includegraphics[width=4cm]{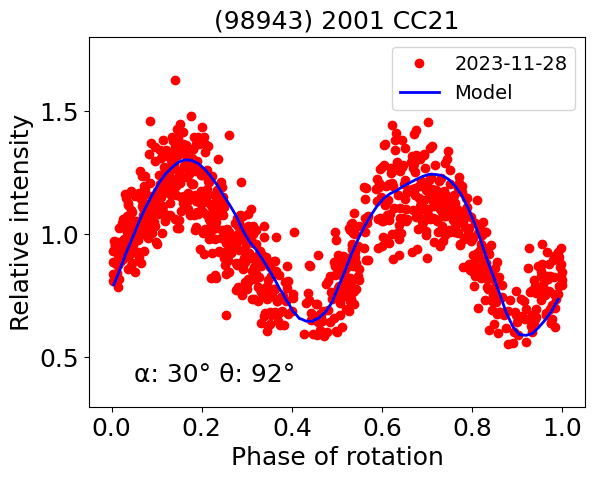}
        \includegraphics[width=4cm]{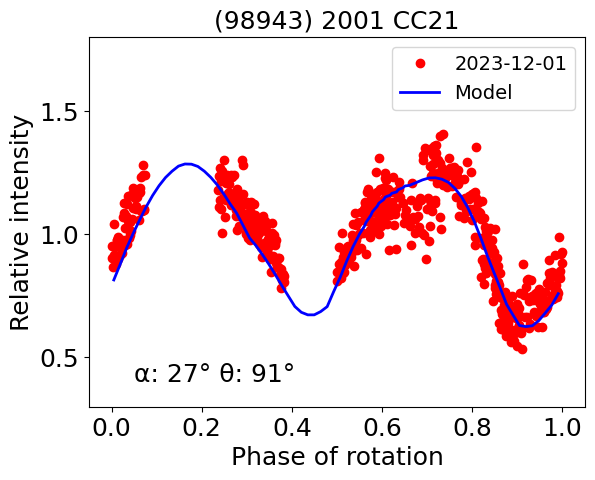}
        \includegraphics[width=4cm]{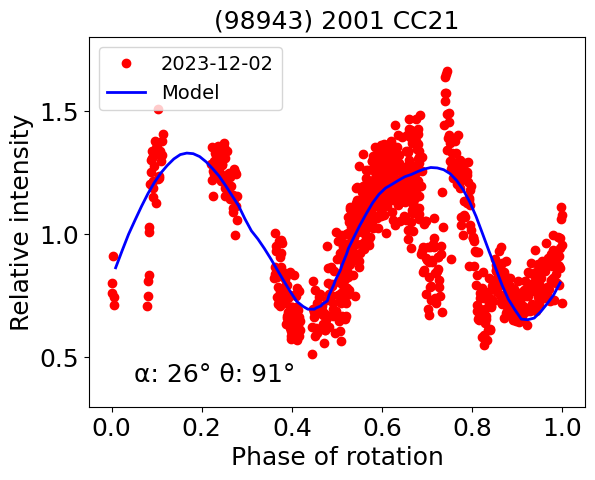}
        \includegraphics[width=4cm]{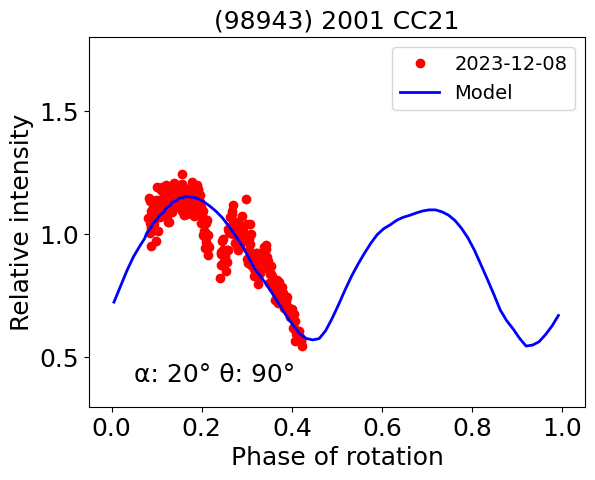}
        \includegraphics[width=4cm]{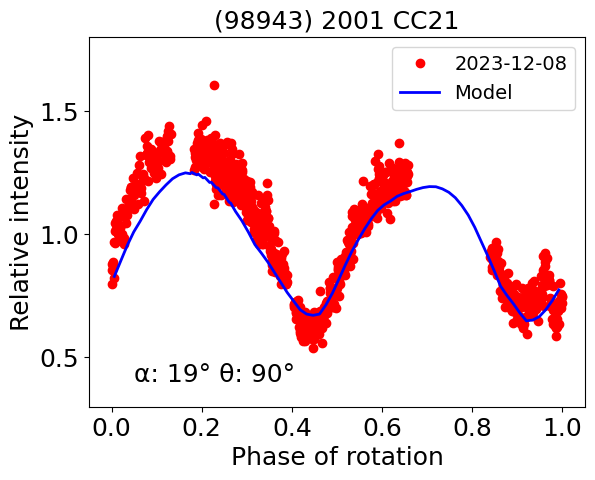}
        \includegraphics[width=4cm]{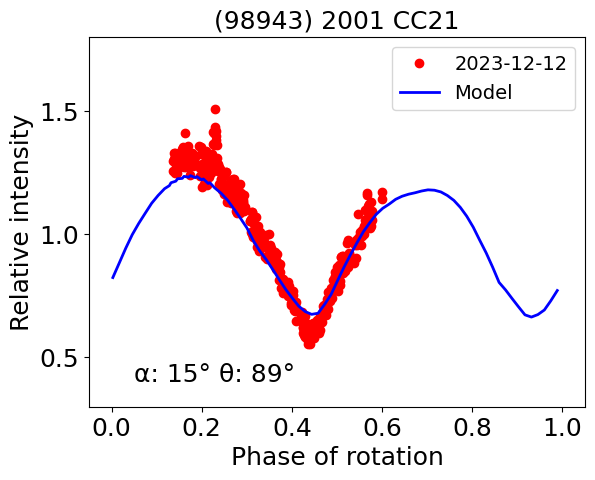}
        \includegraphics[width=4cm]{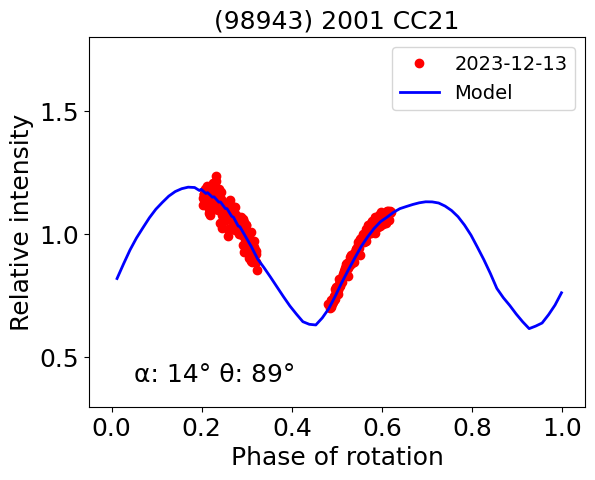}
        \includegraphics[width=4cm]{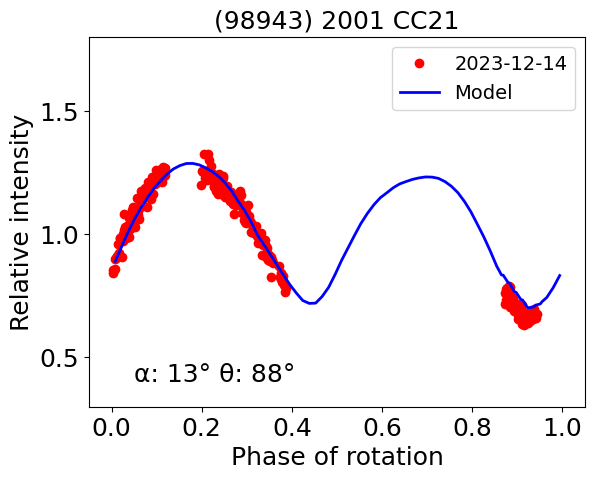}
        \includegraphics[width=4cm]{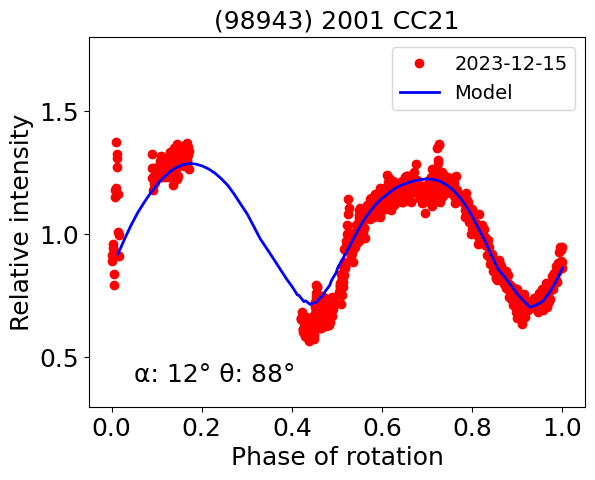}
        \includegraphics[width=4cm]{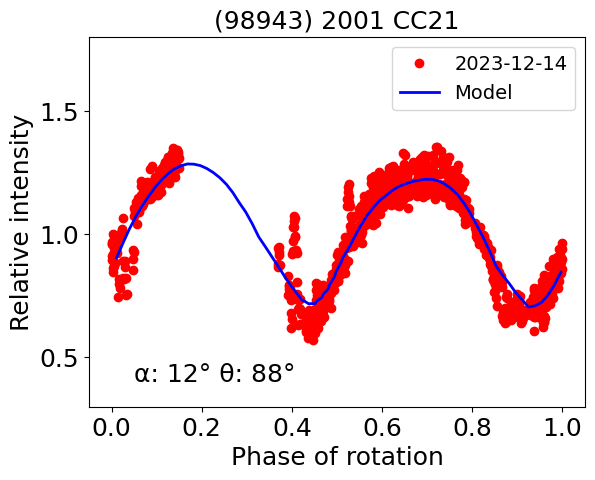}
        \includegraphics[width=4cm]{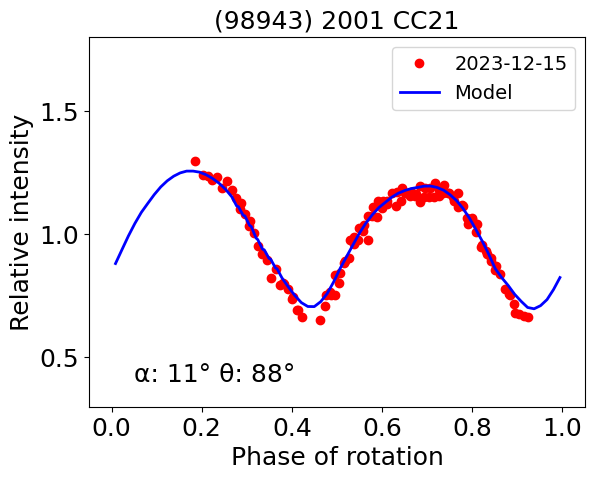}
        \includegraphics[width=4cm]{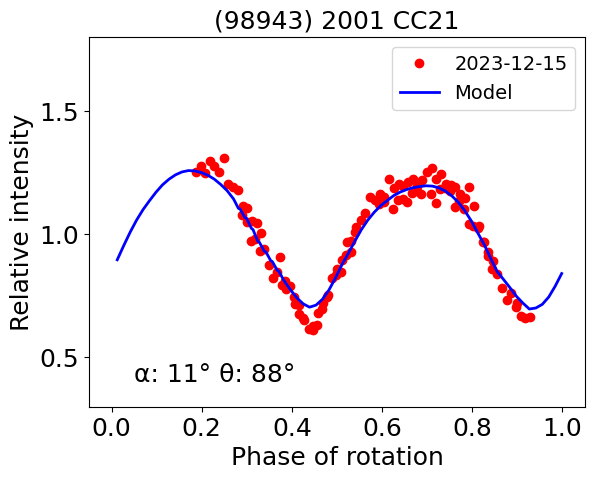}
        \includegraphics[width=4cm]{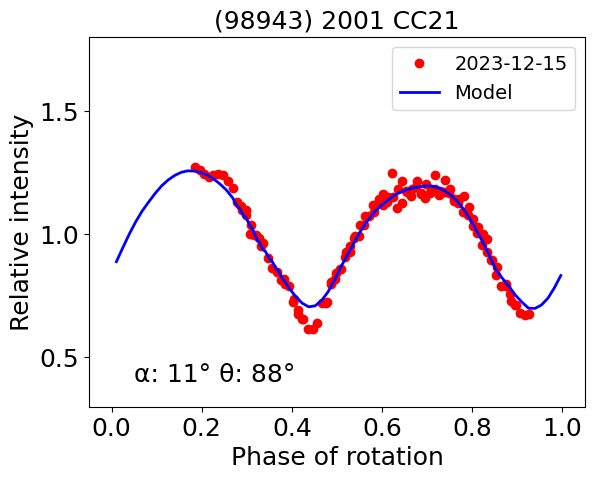}
        \includegraphics[width=4cm]{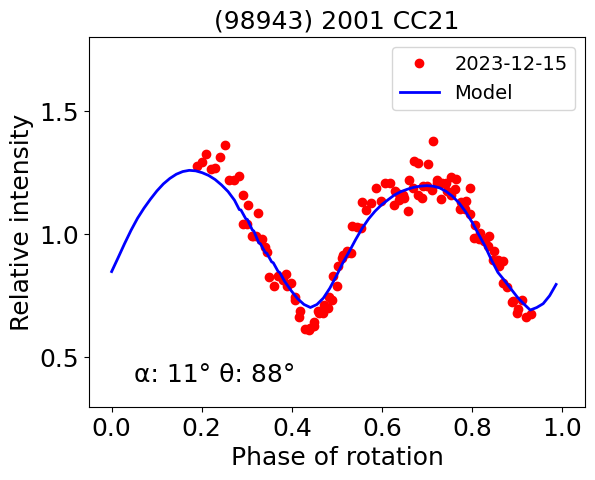}
        \includegraphics[width=4cm]{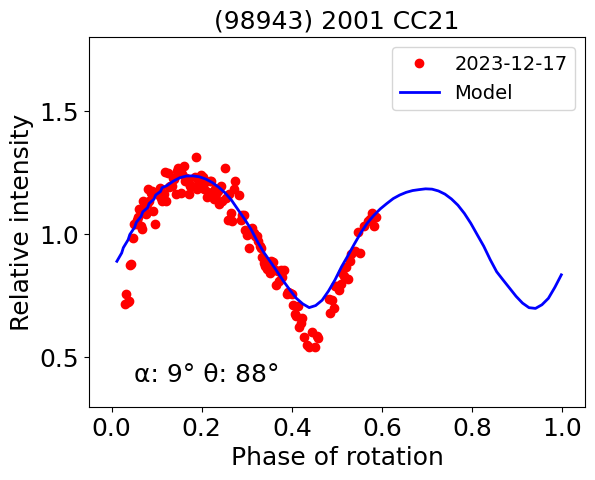}
        \includegraphics[width=4cm]{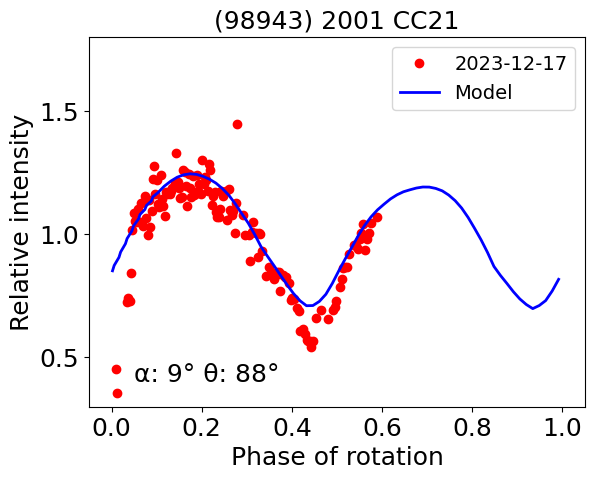}
        \includegraphics[width=4cm]{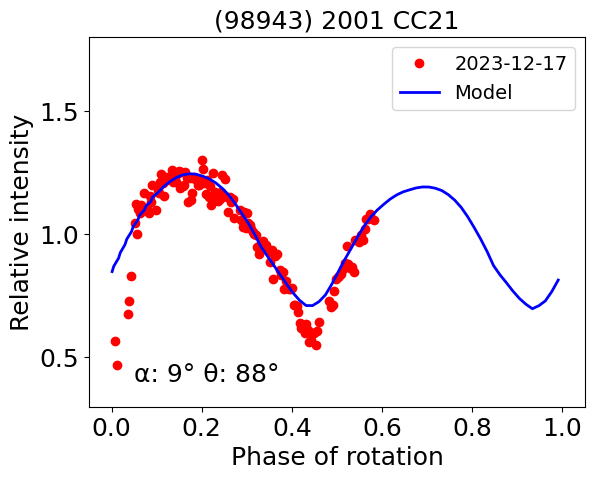}
        \includegraphics[width=4cm]{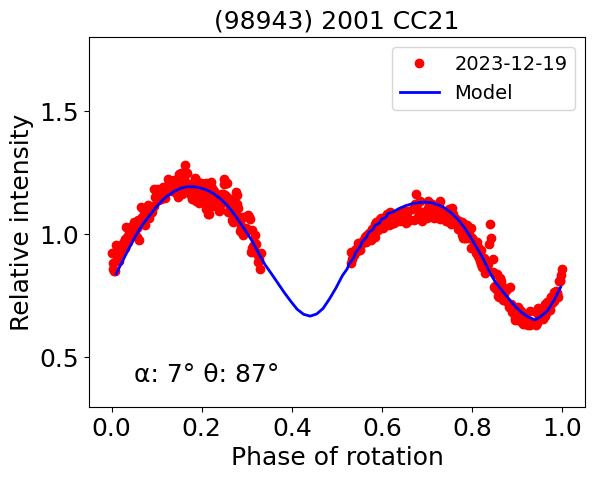}
        \includegraphics[width=4cm]{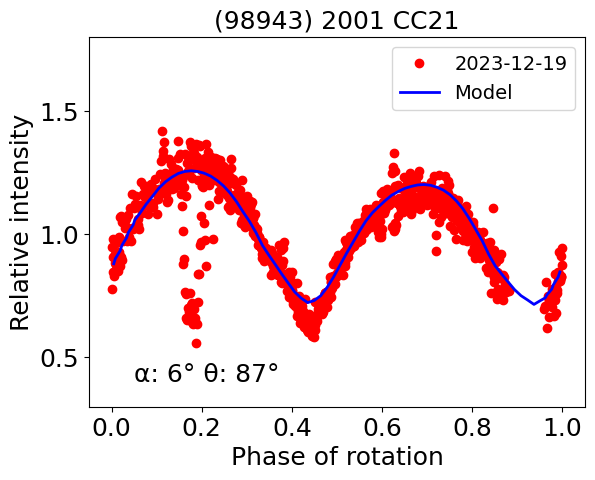}
    \caption{(Continued).}
\end{figure*}

\addtocounter{figure}{-1}
\begin{figure*}
\centering
        \includegraphics[width=4cm]{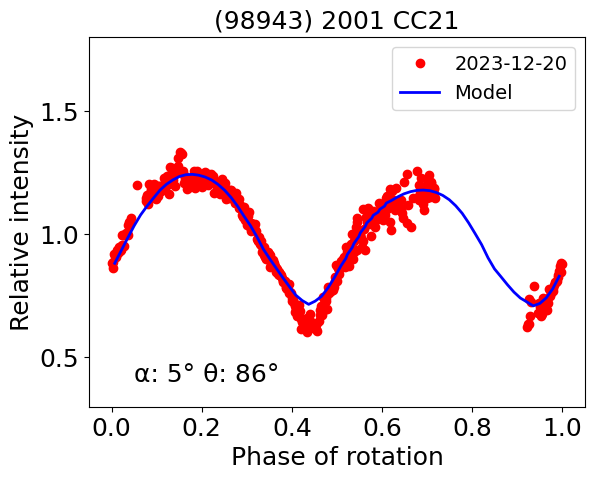}
        \includegraphics[width=4cm]{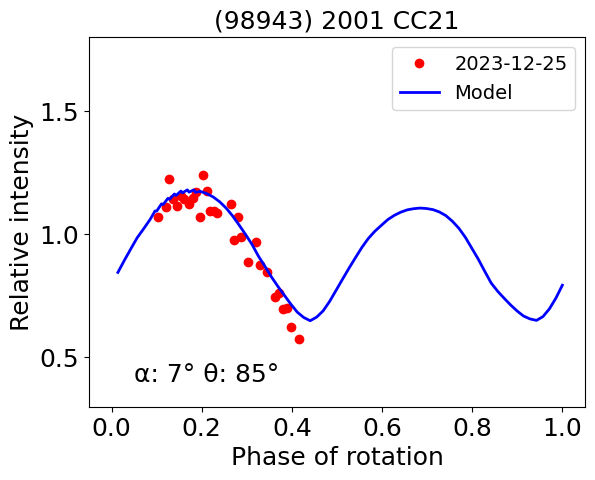}
        \includegraphics[width=4cm]{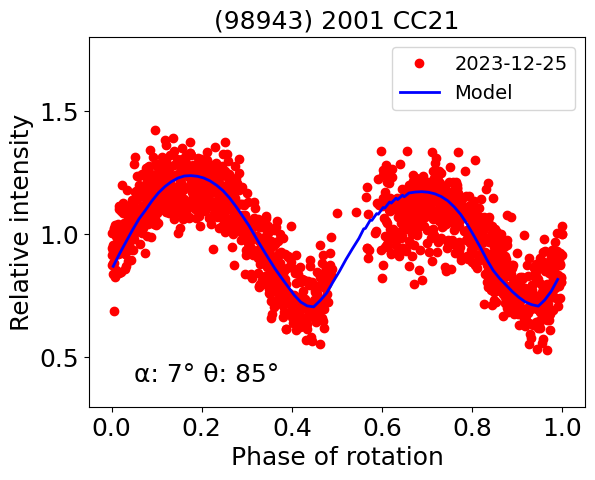}
        \includegraphics[width=4cm]{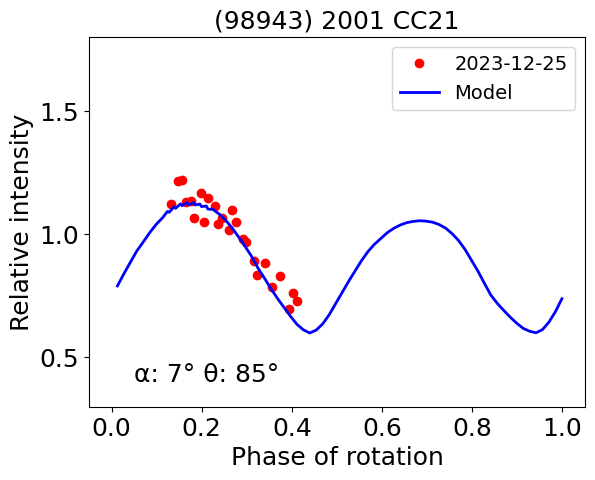}
        \includegraphics[width=4cm]{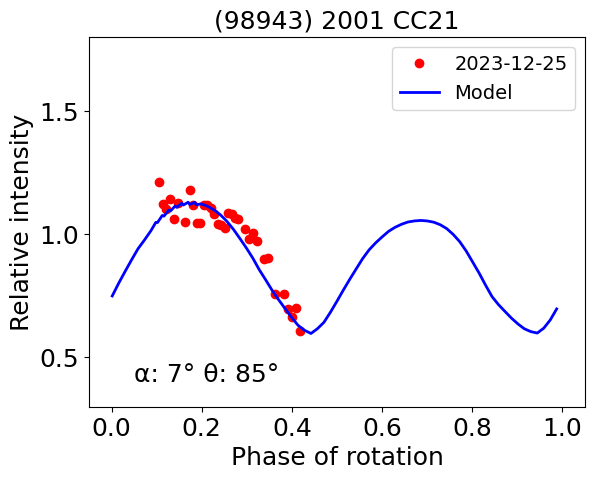}
        \includegraphics[width=4cm]{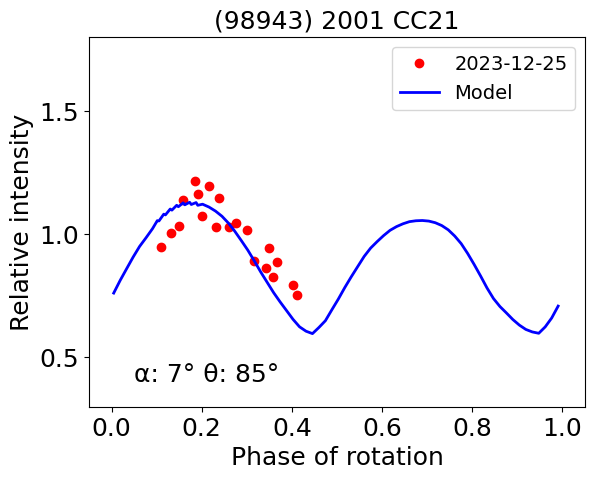}
        \includegraphics[width=4cm]{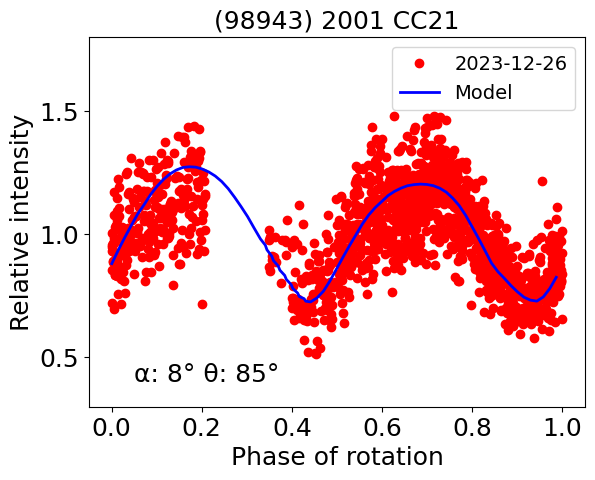}
        \includegraphics[width=4cm]{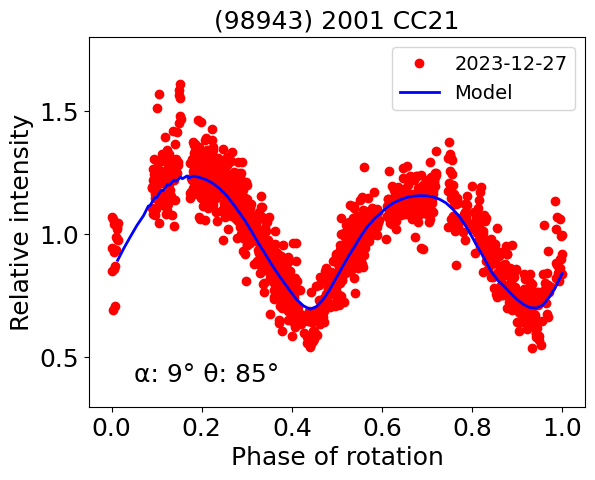}
        \includegraphics[width=4cm]{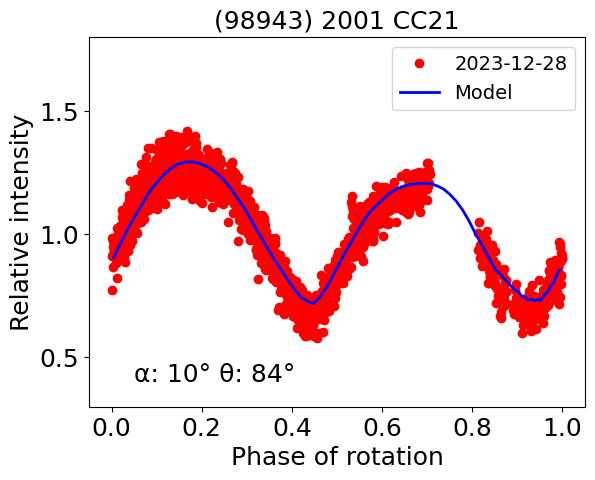}
        \includegraphics[width=4cm]{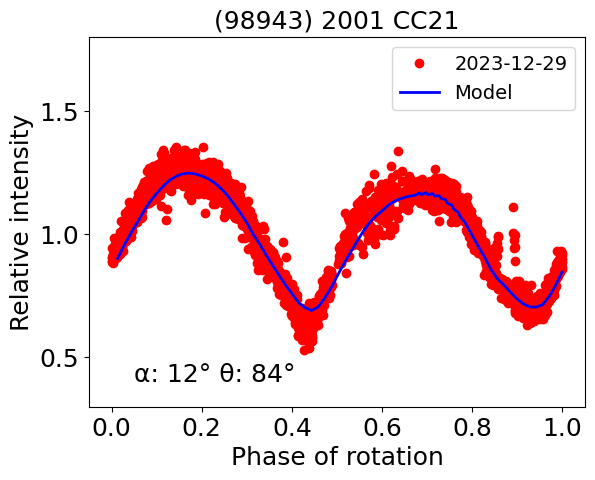}
        \includegraphics[width=4cm]{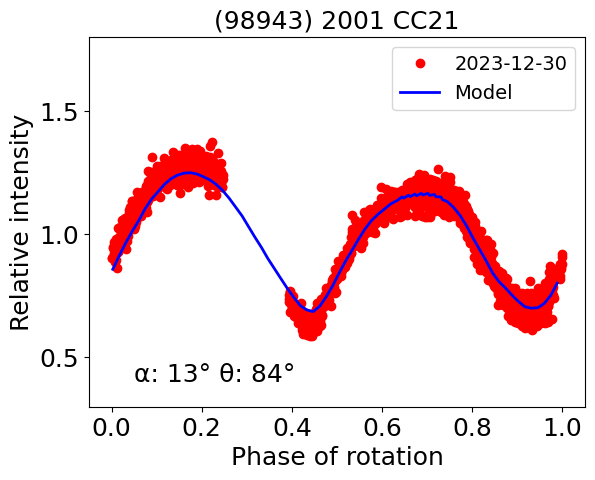}
        \includegraphics[width=4cm]{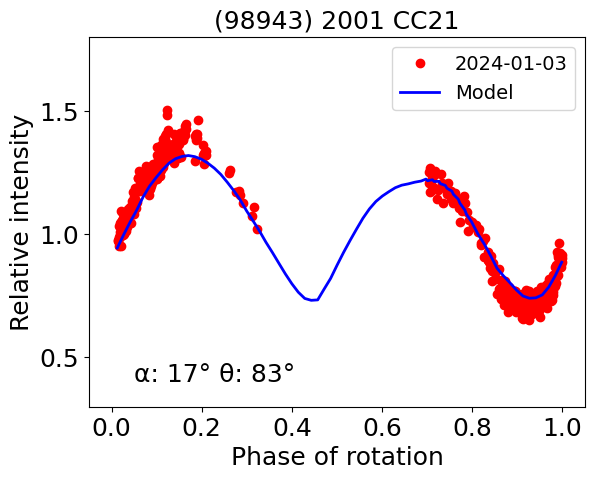}
        \includegraphics[width=4cm]{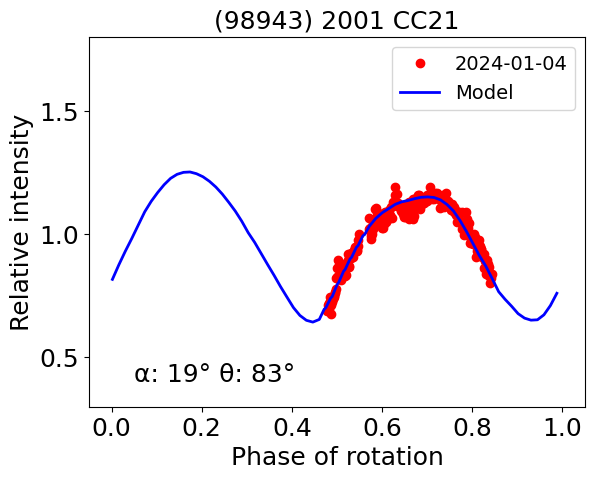}
        \includegraphics[width=4cm]{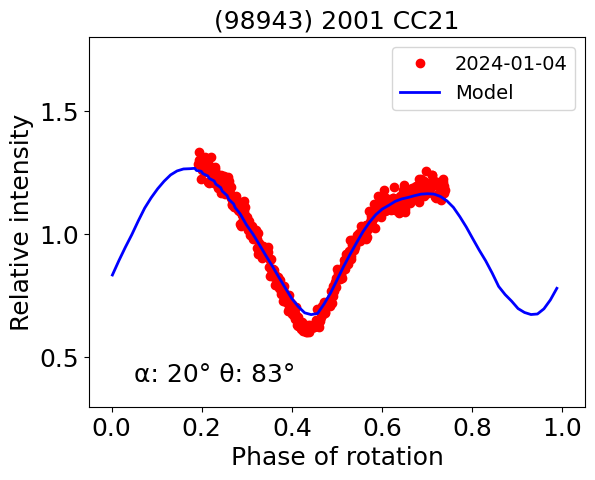}
        \includegraphics[width=4cm]{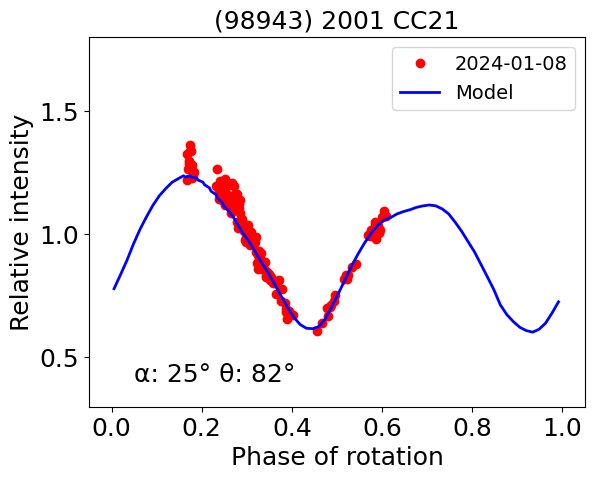}
        \includegraphics[width=4cm]{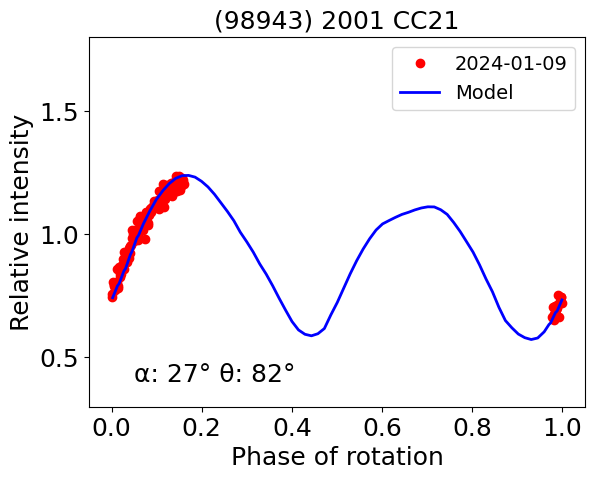}
        \includegraphics[width=4cm]{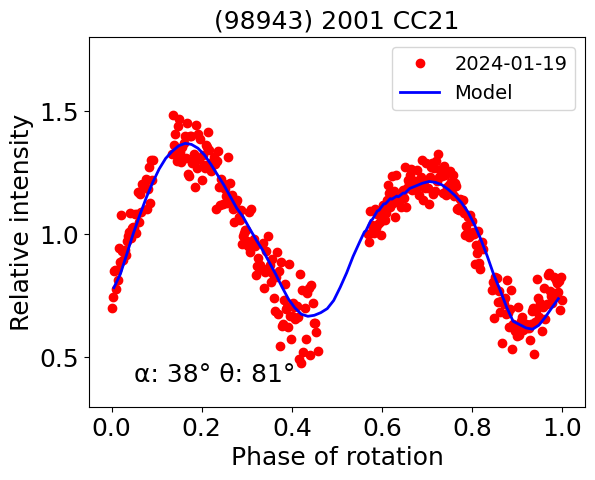}
        \includegraphics[width=4cm]{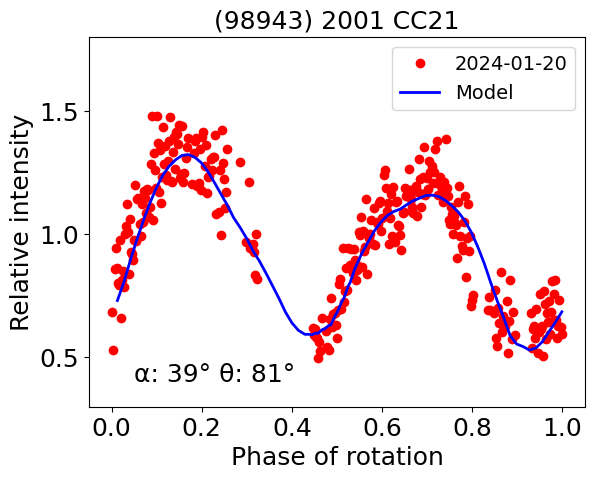}
    \caption{(Continued).}
\end{figure*}

\section{Spectral Comparison}

\begin{figure*}
\centering
        \includegraphics[width=12cm]{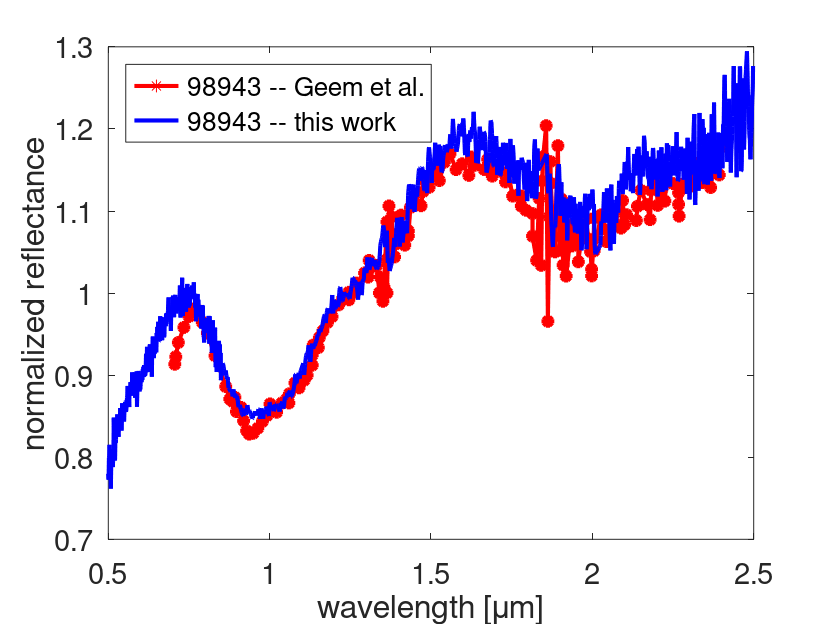}
    \caption{ Comparison between the spectrum of (98943) Torifune reported in this work and the one presented by \citet{2023MNRAS.525L..17G}. Both spectra were normalized at 1.25 $\mu m$.}
    \label{fig:98943oursvsGeem}
\end{figure*}

\bibliography{98943}{}
\bibliographystyle{aasjournal}

%% This command is needed to show the entire author+affiliation list when
%% the collaboration and author truncation commands are used.  It has to
%% go at the end of the manuscript.
%\allauthors

%% Include this line if you are using the \added, \replaced, \deleted
%% commands to see a summary list of all changes at the end of the article.
%\listofchanges

\end{document}